\documentclass[9pt,twocolumn,twoside]{osajnl}

\journal{josaa} 

\setboolean{shortarticle}{false} 

\usepackage{color}
\usepackage{threeparttable}
\usepackage{amsmath}
\usepackage{color}
\usepackage{graphicx}
\usepackage{amsmath}

\newcommand{\be}{\begin{equation}}
\newcommand{\ee}{\end{equation}}
\newcommand{\bea}{\begin{eqnarray}}
\newcommand{\eea}{\end{eqnarray}}

\newcommand{\p}{\partial}

\newcommand{\la}{\langle}
\newcommand{\ra}{\rangle}
\newcommand{\lb}{\left[}
\newcommand{\rb}{\right]}
\newcommand{\lp}{\left(}
\newcommand{\rp}{\right)}
\renewcommand{\vec}[1]{{\bf #1}}
\def\nn{\nonumber\\}

\title{High-resolution 3D phase imaging using a partitioned detection aperture: a wave-optic analysis}

\author[1,*]{Roman Barankov}
\author[1]{Jean-Charles Baritaux}
\author[1]{Jerome Mertz}

\affil[1]{Department of Biomedical Engineering, Boston University\\ 44 Cummington Mall, Boston, Massachusetts 02215, USA}

\affil[*]{Corresponding author: barankov@bu.edu}

\dates{Compiled October 18, 2015}

\ociscodes{(120.5050) Phase measurement; (350.5030) Phase; (110.1220) Apertures; (010.7350) Wave-front sensing.}

\doi{\url{http://dx.doi.org/10.1364/ao.XX.XXXXXX}}

\begin{abstract} 

Quantitative phase imaging has become a topic of considerable interest in the microscopy community. We have recently described one such technique based on the use of a partitioned detection aperture, which can be operated in a single shot with an extended source [Opt. Lett. {\bf 37}, 4062 (2012)]. We follow up on this work by providing a rigorous theory of our technique using paraxial wave optics, where we derive fully three-dimensional spread functions for both phase and intensity. Using these functions we discuss methods of phase reconstruction for in- and out-of-focus samples, insensitive to weak attenuations of light. Our approach provides a strategy for detection-limited lateral resolution with an extended depth of field, and is applicable to imaging smooth and rough samples.

\end{abstract}

\setboolean{displaycopyright}{true}

\begin{document}

\maketitle
\thispagestyle{fancy}
\ifthenelse{\boolean{shortarticle}}{\abscontent}{}


\section{Introduction}

Quantitative phase imaging reveals optical path variations in almost
transparent (and also specularly reflecting) objects, and thus provides useful
information for biological or industrial applications. Different strategies
exist to obtain phase information \cite{Popescu_book}, most of which are
interferometric and rely on monochromatic illumination. In the case of
quasi-monochromatic illumination, optical phase is not well defined and
interferometric techniques only measure changes in optical phase relative to a
self-reference provided, for example, by spatial
filtering~\cite{Zernike1935,Popescu2004,Bernet2006}, or
shearing~\cite{Nomarski1955,Arnison2004,Shribak2008}. Alternatively, gradients
in optical phase can be inferred non-interferometrically based purely on
intensity imaging, using, for example, the transport of intensity equation
\cite{Streibl1984,Paganin1998,Kou2010}, or by directly measuring changes in
the direction of flux density, or local light
tilt~\cite{Stewart1976,Platt2001,Yi2006,Mehta2009,Bon2009,Iglesias2011,Tian2014}.
We have recently introduced a method to measure local light tilts that is
based on the use of a partitioned detection aperture~\cite{Mertz2012}. By
associating local light tilts with phase gradients and integrating these over
space, we thus obtain quantitative images of phase. The method is passive,
single-shot, provides high spatial resolution, and works with quasi-broadband,
partially coherent illumination. A reflection version of the method is
described in~\cite{Barankov2013}, as is a version implemented for closed-loop
adaptive optic wavefront sensing~\cite{Li2015}.

The purpose of this work is to present an in depth theoretical analysis of
our partitioned aperture wavefront imaging technique that goes significantly
beyond the cursory and mostly experimental expositions we provided in our previous
reports~\cite{Mertz2012,Barankov2013}. In particular, our previous reports
were based on smooth phase approximations and on strict assumptions regarding
the numerical apertures of both the illumination and detection optics of our
system, which ultimately limited spatial resolution. In this paper, we
generalize to phases that are not necessarily smooth (though they must remain
small), and we relax our assumptions regarding numerical apertures, enabling
access to improved spatial resolution. Finally, we extend our analysis to
three dimensions by considering samples that can be out of focus, and
describing a refocusing strategy conceptually similar to that used in light
field~\cite{Ng2005,Levoy2006} or integral~\cite{Xiao2013} imaging. The goal of
this paper is to provide a rigorous theoretical underpinning to partitioned
aperture wavefront imaging, which has been long overdue. In the process, we also discuss strategies to expand on its capabilities.

\section{Phase imaging with a partitioned aperture}

\begin{figure}[htbp]
	\centering
    \includegraphics[width=8cm]{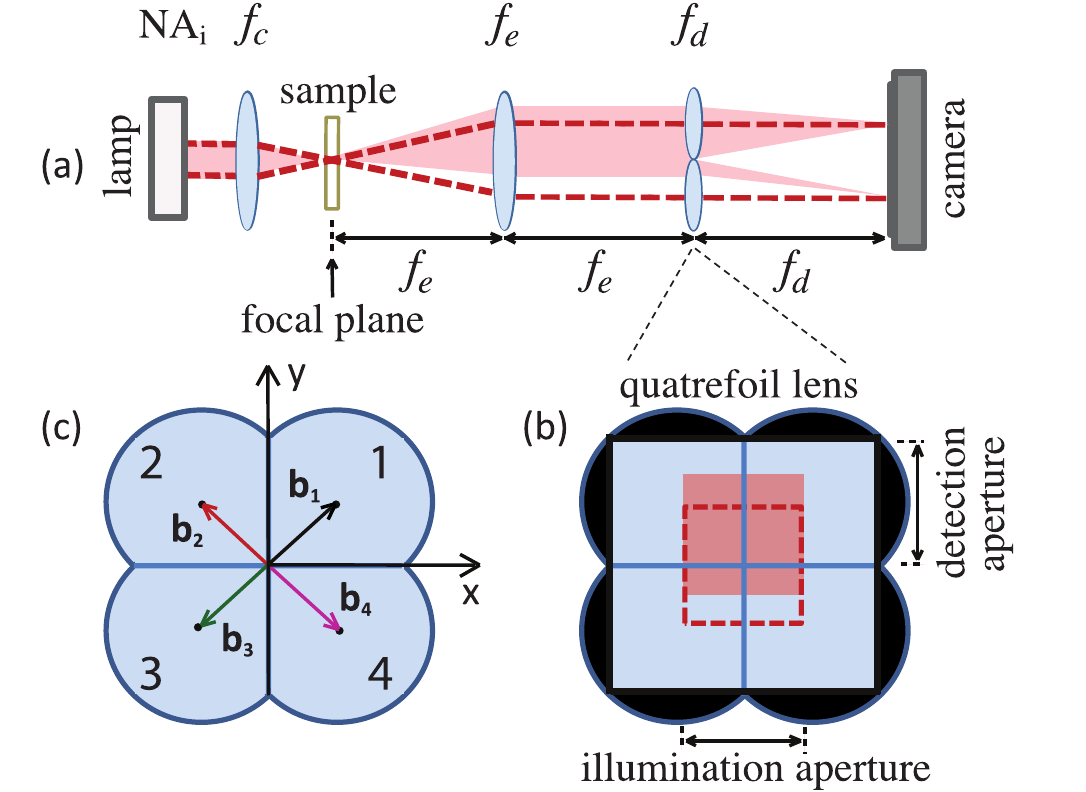}
     \caption{(a) Partitioned detection aperture microscope in transmission mode. (b) Image of the illumination aperture on the face of the partitioned assembly of four lenses: dashed square identifies the aperture in the absence of a phase sample, the shifted solid square demonstrates the effect of light refraction by the sample. (c) Four off-axis lens assembly: the locations of optical axes of the four lenses with respect to the optical axis of the system are identified by vectors $\vec b_i$, $i=1..4$. A larger aperture (shading) ensures the detection apertures are square.}
\label{fig:setup}
\end{figure}

A schematic of our method is shown in Fig.~\ref{fig:setup}. A phase sample is
modeled as a thin transparent layer of variable optical thickness. The sample
is trans-illuminated by an axially symmetric beam of light of uniform
intensity, which is characterized at each point by a distribution of light
rays filling a well-defined square illumination numerical aperture. After
traversing the sample, the light is imaged by a device that comprises a lens
of focal length $f_e$, a partitioned lens assembly of focal length $f_d$ and a
camera, arranged in a $3f$ imaging configuration. For simplicity, we assume
that the lens assembly is masked such that each individual lens in the
assembly appears square.

The purpose of the quatrefoil lens assembly is to simultaneously project
multiple separated images onto the camera (here four), each acquired from a
different oblique direction. Local tilts of light rays caused by the sample
thus lead to local intensity differences in the projected images, as can be
understood from Fig.~\ref{fig:setup}. In the absence of a sample, the square
illumination aperture forms a centered image on the surface of the lens
assembly, and one obtains four images of uniform and equal intensities at the
camera. When a phase sample is present, the intensity distribution between the
images changes. Indeed, phase slopes at a given sample point cause the light
rays through that point to tilt and the image of the illumination aperture
from that point to shift (Fig.~\ref{fig:setup}b). Correspondingly, the four
images of the sample point projected onto the camera exhibit unequal
intensities. If the detection aperture is large enough there is no vignetting
of the illumination aperture image and one can uniquely retrieve the tilts
from the intensity distribution of the four images. As we discuss below, the
retrieved light tilts provide a measure of sample-induced phase gradients, and
hence, ultimately, of phase.

\subsection{Partitioned aperture imaging}

The four off-axis lenses form four images at the detector plane. To begin, we
consider light propagating through one of the lenses shifted from the optical
axis of the system by vector $\vec b$. Throughout this work we confine
ourselves to the paraxial wave approximation~\cite{MertzBook} and assume, for
simplicity, unit magnification (i.e. $f_e=f_d$). The field at the detector
(camera) plane is thus given by 
\bea
E_d(\vec r,t)&=&e^{i6\pi k f_e}e^{i\pi\frac{k}{f_e}\vec r^2}\int {\rm d^2} \vec r_0 e^{-i2\pi\frac{k}{f_e}\vec r_0\vec b}\nn
&& \times\,{\rm CSF_d}\lp\vec r-\vec r_0\rp E_0(\vec r_0,t),
\eea
where $\lambda=1/k$ is the wavelength, the coordinate $\vec r=\vec r_i-\vec b$
accounts for the shift of the image with respect to the position $\vec b$ of
the detection lens, and $E_0(\vec r_0,t)$ is the illumination field at time
$t$ at the focal plane (in the absence of a sample). The detection numerical
aperture of the system is defined by that of an individual off-axis detection
lens $f_d$, and the coherent spread function of each lens, when centered, is
given by
\be
{\rm CSF_d}(\vec r)=\lp\frac{k}{f_e}\rp^2\int {\rm d^2} \vec r'\,e^{i2\pi\frac{k}{f_e}\vec r'\vec r}{\rm A_d}(\vec r')
\ee
where ${\rm A_d}(\vec r)$ is the square aperture function of an individual
lens.

The intensity at the detector plane is defined as $I_d(\vec r)=\la E_d(\vec
r,t) E^*_d(\vec r,t)\ra$, where the average $\la...\ra$ is taken over a time
scale much longer than the coherence time of the light. The intensity is
expressed as a weighted sum of the mutual intensity $J_0(\vec r_{0c},\vec
r_{0d})$ over the focal plane
\bea\label{eq:intensity_image_general}
I_d(\vec r)&=&\int {\rm d^2} \vec r_{0c}{\rm d^2} \vec r_{0d}\,e^{-i2\pi\frac{k}{f_e}\vec b \vec r_{0d}} J_0(\vec r_{0c},\vec r_{0d})\nn
&&\times\,{\rm CSF_d}\lp\vec r-\vec r_{0+}\rp{\rm CSF^*_d} \lp\vec r-\vec r_{0-}\rp ,
\eea
where $J_{0}(\vec r_{0c},\vec r_{0d})=\la E_0(\vec r_{0+},t) E^*_0(\vec
r_{0-},t)\ra$ characterizes spatial correlations of the illumination field
between pairs of points in the focal plane, and $\vec r_{0\pm}=\vec
r_{0c}\pm\vec r_{0d}/2$. The off-axis shift of the detection lens is accounted
for by the phase factor with the spatial frequency proportional to the oblique
detection tilt angle $\vec b/f_e$. Eq.~(\ref{eq:intensity_image_general})
serves as the basis for what follows.

\subsection{Partially coherent illumination}

In our optical system a transparent sample is trans-illuminated by a uniform
beam of quasi-monochromatic light resulting from K\"ohler illumination with a
condenser lens of focal length $f_c$. The spatial extent of the light source
is defined by the aperture function ${\rm A_c}(\vec r)$. We consider a
spatially incoherent source described by the function $J_{c}(\vec r_{c},\vec
r_{d})= k^{-2}I_0\, {\rm A_c}(\vec r_{c})\delta(\vec r_{d})$, where $I_0$ is
the uniform intensity of the source, and $\delta(\vec r)$ is the
two-dimensional Dirac delta function. The illumination mutual intensity at the
focal plane (in the absence of a sample) is then given by $J_0(\vec r_{c},\vec
r_{d})=I_0\mu_0(\vec r_{d})$, where the illumination coherence function is
given by
\be\label{eq:mu_square}
\mu_0(\vec r)=\frac{1}{f_c^2\Omega_c}\int {\rm d^2} \vec r' e^{-i2\pi\frac{k}{f_c}\vec r'\vec r}{\rm A_c}(\vec r'),
\ee
and we defined the solid angle $\Omega_c=\int {\rm d^2} \vec r\left|{\rm
A_c}(\vec r')\right|^2/f_c^2$ of the source as viewed through the condenser
lens. As noted above, we assume a square illumination aperture of size
$2d_c\times 2d_c$. The illumination coherence function is thus separable
$\mu_0(\vec r)={\rm sinc}\lp \pi x/a_{i}\rp{\rm sinc}\lp \pi y/a_{i}\rp$,
where $a_{i}=\lambda/(2{\rm NA_i})$ defines a characteristic length scale of
the illumination mutual intensity, ${\rm NA_i}=d_{c}/f_c$ is the numerical
aperture of the illumination, and ${\rm sinc}(x)=\sin (x)/x$. The solid angle
of the source is $\Omega_c=(2{\rm NA_i})^2$.

\subsection{Thin phase sample}
A thin phase sample is characterized by a transmissivity function $T(\vec
r)=e^{i\phi(\vec r)}$. The complex-valued phase of the sample $\phi(\vec
r)=\varphi(\vec r)+i\alpha(\vec r)$ accounts for spatial variations in the
optical path and attenuation of light fields in the sample. The phase can be
approximated by $\phi(\vec r)=(2\pi/\lambda) \int_0^{d} dz\, n(\vec r,z)$,
where $n(\vec r,z)$ is the local index of refraction, and the thickness
$d(\vec r)$ of the sample is assumed small. In the reflection mode, the real
part of the phase is related to the topographic profile of the
sample~\cite{Barankov2013}. The field after propagation through the sample is
$E(\vec r,t)T(\vec r)=E(\vec r,t)e^{i\phi(\vec r)}$. Correspondingly, the
mutual intensity immediately after the sample depends on the relative phases
between pairs of points
\be\label{eq:mutual_coherence_sample}
J_s(\vec r_{c},\vec r_{d})=J(\vec r_{c},\vec r_{d}) \exp\lb i\phi\lp\vec r_{c}+\frac{\vec r_{d}}{2}\rp-i\phi^*\lp\vec r_{c}-\frac{\vec r_{d}}{2}\rp\rb,
\ee
where $J(\vec r_c,\vec r_d)$ is the mutual intensity of the illumination field
$E(\vec r,t)$ immediately before the sample.

\section{Approximation of smooth phase gradients}

The in-focus imaging of smooth phase gradients has been previously described
in Ref.~\cite{Stewart1976, Iglesias2011,Mertz2012} using geometrical optics.
In this section we provide an alternative derivation of the formalism based on
the wave optics, and identify its limitations.

When we employ the smooth-phase approximation $\phi(\vec r_{c}\pm \vec
r_{d}/2)\approx \phi(\vec r_{c})\pm (\vec r_{d}/2)\nabla \phi(\vec r_{c})$ ,
the mutual intensity ~(\ref{eq:mutual_coherence_sample}) at the focal plane
immediately after the sample becomes
\be\label{eq:mutual_coherence_gradient}
J_0(\vec r_{0c},\vec r_{0d})\approx I_0 \mu_0(\vec r_{0d}) \exp\lb i \vec r_{0d}\nabla\varphi( \vec r_{0c})-2\alpha(\vec r_{0c})\rb.
\ee
In this manner, sample-induced phase gradients become imprinted on the optical
mutual intensity, provided the coherence function $\mu_0(\vec r_{d})$ is at
least partially coherent.

The approximation~(\ref{eq:mutual_coherence_gradient}) is valid provided
$a_i^2|\nabla^3\varphi|\ll |\nabla\varphi|$ and $a_i^2|\nabla^2\alpha|\ll \alpha$, that is provided phases are smooth.
When more stringent conditions $a_{i}|\nabla^2\varphi|\ll |\nabla\varphi|$ and $a_i|\nabla \alpha|\ll\alpha$ are
satisfied, that is when the phase gradients are smooth, we can approximate
$\nabla\varphi(\vec r_{0c})\approx\nabla\varphi(\vec r_{c})$ and $\alpha(\vec
r_{0c})\approx \alpha(\vec r_c)$ in Eq.~(\ref{eq:mutual_coherence_gradient}).
Substituting the mutual intensity in Eq.~(\ref{eq:intensity_image_general}),
and also using the relation~(\ref{eq:mu_square}) between the Fourier
components of the mutual intensity and the aperture function of the condenser (see Appendix for our convention for Fourier transforms),
$\tilde\mu_0 (\vec k)={\rm A_c}\lp -f_c\vec k/k\rp/(k^2\Omega_c)$, we obtain
the intensity at the detector plane
\bea\label{eq:intensity_image_smooth}
I_d(\vec r)&\approx& I_0\,e^{-2\alpha(\vec r)}\frac{1}{f_c^2\Omega_c}\int {\rm d^2} \vec r' \left |{\rm A_d}\lp\frac{f_e}{f_c}\vec r'-\vec b\rp\right|^2\nn
&& \times\,{\rm A_c}\lp \frac{f_c}{2\pi k}\nabla\varphi(\vec r)-\vec r'\rp.
\eea
This equation can be readily interpreted from geometrical optics. According to
Eq.~(\ref{eq:intensity_image_smooth}), light rays propagating within the cone
limited by the illumination numerical aperture are tilted by a common angle at
a given sample point. This tilt is translated into an effective lateral shift
of the illumination aperture that is proportional to the tilt. The light is
collected by the detection aperture ${\rm A_d}$, itself offset from the
optical axis by vector $\vec b$. Finally, the image is formed as an incoherent
sum of the rays. The intensity is reduced exponentially according to the local
value of attenuation parameter $\alpha(\vec r)$.

\subsection{Imaging smooth phase gradients using a quatrefoil partitioned
aperture}

The integration in Eq.~(\ref{eq:intensity_image_smooth}) can be carried out
explicitly when the illumination aperture ${\rm A_c}(\vec r)$ is a square. We
assume that the coordinate system is centered at the optical axis of the
system, and that the axes are aligned with the sides of the aperture. In a
quatrefoil assembly, the centers of four lenses are placed symmetrically at
positions
\bea\label{eq:quatre-foil lens}
\vec b_1=(b,b),\,\vec b_2=(-b,b),\,\vec b_3=(-b,-b),\,\vec b_4=(b,-b),
\eea 
where $b$ is the distance of a lens center from the two axes $(x,y)$ (see
Fig.~\ref{fig:setup}c). The four lenses of the assembly collect different
amounts of light depending on the local tilt angles imparted on the light by
the sample, and form images of varying intensity.

Assuming the detection aperture of each lens exceeds the illumination
aperture, from Eq.~(\ref{eq:intensity_image_smooth}) we obtain
\be\label{eq:int_smooth_grad}
I_d(\vec r)=\frac{I_0}{4}\,e^{-2\alpha(\vec r)}\lb 1\pm\frac{\theta_x(\vec r)}{{\rm NA_i}}\rb\lb 1\pm\frac{\theta_y(\vec r)}{{\rm NA_i}}\rb,
\ee
where we have introduced the local tilt angles $\theta_x$ and $\theta_y$ along
$x$ and $y$ axes, respectively. The connection between these local tilt angles
of light and the local phase gradients of the sample is given by
\be\label{eq:phase_gradients}
\theta_x=\frac{\lambda}{2\pi}\p_x\varphi,\quad \theta_y=\frac{\lambda}{2\pi}\p_y\varphi,
\ee
which constitutes the general crux of intensity-based phase imaging techniques
\cite{Nugent_review}. We note that the signs of tilt angles in
Eq.~(\ref{eq:int_smooth_grad}) are defined by the quadrant of the detection
lens. For example, if the lens occupies the fourth quadrant as seen from the
incoming beam, that is $\vec b=(+b,-b)$, one should take the signs $(+)$ and
$(-)$ for the $x$ and $y$ components of the tilts, respectively. The factor
$1/4$ in Eq.~(\ref{eq:int_smooth_grad}) accounts for the splitting of the lens
assembly into four equal parts.

The light tilt angles are extracted from the four images at the camera using
the linear combinations:
\bea\label{eq:tilts}
\theta_x&=&{\rm NA_i}\lp I_{1}+I_{4}-I_{2}-I_{3}\rp/I_{tot},\nn
\theta_y&=&{\rm NA_i}\lp I_{1}+I_{2}-I_{3}-I_{4}\rp/I_{tot},
\eea
where $I_k$ for $k=1..4$ are the intensities detected in the four quadrants of
the detector plane, and the total intensity is $I_{tot}=\sum_{k=1..4} I_k$.
According to Eq.~(\ref{eq:int_smooth_grad}), the range for light tilt
measurements is defined by the illumination numerical aperture, such that
$|\theta_{x,y}|\le {\rm NA_i}$. The tilt angles can be expressed using the
wavevector components, $\theta_{x,y}= k_{x,y}/k$, so that the range of
the tilts can be also written as $|k_{x,y}|\le 1/(2a_i)$.
Eqs.~(\ref{eq:phase_gradients}) and~(\ref{eq:tilts}) have been previously
derived in Ref.~\cite{Stewart1976,Iglesias2011,Mertz2012} using geometrical
optics.

Eqs.~(\ref{eq:phase_gradients}) and~(\ref{eq:tilts}) can be used to
reconstruct the sample phase distribution. 
Assuming that the spatial support of the sample phase $\varphi(\vec r)$ is
finite within the imaging field of view, we find the Fourier components of the
tilts~(\ref{eq:tilts}) and use Eq.~(\ref{eq:phase_gradients}) to obtain the
spectral representation of phase~\cite{Arnison2004} $\tilde\varphi(\vec
k)=-ik\lb\tilde\theta_x(\vec k)+i\tilde\theta_y(\vec k)\rb/(k_x+ik_y)$. The
phase profile is found by using the inverse Fourier transformation of
$\tilde\varphi(\vec k)$
\be\label{eq:reconstruct_phase_smooth}
\varphi(\vec r)=-ik\int {\rm d^2} \vec k\, e^{i2\pi \vec k\vec r}\,\frac{\tilde\theta_x(\vec k)+i\tilde\theta_y(\vec k)}{k_x+ik_y},
\ee
where the integration extends over the bandwidth of optical system limited by
the detection aperture. The reconstruction~(\ref{eq:reconstruct_phase_smooth})
was derived in~\cite{Mertz2012} without mention of the spatial resolution of
the method. Since the derivation assumed that phase gradients are smooth on
the spatial scale of the mutual intensity, we conclude
that~(\ref{eq:reconstruct_phase_smooth}) is accurate for spatial frequencies
of the phase much smaller than the tilt range, $|k_{x,y}|\ll
1/(2a_i)$. In the following section we derive an alternative phase
reconstruction method characterized instead by a detection-limited spatial
resolution.

\section{Approximation of small phases}
We have derived intensity~(\ref{eq:intensity_image_smooth}) within the
approximation of smooth phase gradients. We now generalize our formalism to
the approximation of small phases, and allow for the possibility of non-smooth
phase distributions. In the process, we also extend our formalism to three
dimensions by allowing the possibility of sample defocus.

We assume that a phase object is located some distance $z$ away from the focal
plane ($z>0$ correspond to displacements away from the imaging lens $f_e$).
Within the approximation of small phases, the mutual
intensity~(\ref{eq:mutual_coherence_sample}) at the sample plane becomes
\be\label{eq:sample_small_phase}
J_s(\vec r_{c},\vec r_{d})\approx J(\vec r_{c},\vec r_{d})\lb 1+i\phi\lp\vec r_{+}\rp-i\phi^*\lp\vec r_{-}\rp\rb,
\ee
where $\vec r_{\pm}=\vec r_{c}\pm\vec r_{d}/2$, and $J(\vec r_{c},\vec r_{d})$
is the mutual intensity immediately before the sample. The small expansion
parameter is the phase difference between two points within the spatial range
of the mutual intensity.

We propagate the illumination mutual intensity from the focal plane toward the
sample by the off-focus distance $z$, where it is modulated according to
Eq.~(\ref{eq:sample_small_phase}). The function is then propagated back to the
focal plane by distance $-z$, where it reads
\bea\label{eq:mutual_coherence_small_phase}
J_0(\vec r_{0c},\vec r_{0d})&\approx & I_0\lp\frac{k}{z}\rp^2\int {\rm d^2}   \vec r_{c}{\rm d^2} \vec r_{d}\,\mu_0(\vec r_{d})
 e^{-i2\pi\frac{k}{z}(\vec r_{c}-\vec r_{0c})(\vec r_{d}-\vec r_{0d})}\nn 
&&\times\lb 1+i\phi\lp\vec r_{+}\rp-i\phi^*\lp\vec r_{-}\rp\rb,
\eea
where $\vec r_{\pm}=\vec r_{c}\pm\vec r_{d}/2$.

Substituting~(\ref{eq:mutual_coherence_small_phase}) into
Eq.~(\ref{eq:intensity_image_general}), we find that the intensity at the
detector plane contains two terms
\be\label{eq:intensity_total}
I_d(\vec r,z)=I_\alpha(\vec r,z)+I_\varphi(\vec r,z).
\ee

The first term $I_\alpha$ accounting for weak attenuation of light in the
sample is given by
\be\label{eq:intensity_attenuation}
I_\alpha(\vec r,z)=I_0\int {\rm d^2}\vec r_0\, P_\alpha(\vec r-\vec r_0,z)\,\lb \frac{1}{2}-\alpha(\vec r_0)\rb,
\ee
where we refer to $P_\alpha(\vec r,z)$ as the intensity spread function. When
attenuation of light is absent ($\alpha=0$), the intensity becomes a constant
independent of out-of-focus position $z$.

The phase term is given by
\be\label{eq:phase_sensitive_intensity}
I_\varphi(\vec r,z)=I_0\int {\rm d^2}\vec r_0\, P_\varphi(\vec r-\vec r_0,z)\,\varphi(\vec r_0),
\ee
where $ P_\varphi(\vec r,z)$ is the phase spread function.

Concise representations of the intensity and phase spread functions are
obtained using Fourier transforms. Assuming finite spatial support of
$\phi(\vec r)$, we find the spatial frequency components of
intensity~(\ref{eq:intensity_total})
\be\label{eq:intensity_phase_fourier}
\tilde I_d(\vec k)/I_0=\lb\delta(\vec k)/2-\tilde\alpha(\vec k)\rb \tilde P_\alpha(\vec k,z)+\tilde P_\varphi(\vec k,z)\tilde\varphi(\vec k),
\ee 
where $\tilde P_\alpha(\vec k,z)$ and $\tilde P_\varphi(\vec k,z)$ are the
intensity and phase transfer functions respectively.

Using the notation $\vec q_+=\vec q +\vec k/2$ and $\vec q_-=\vec q -\vec
k/2$, the intensity transfer function reads
\bea\label{eq:PSF_0}
\tilde P_\alpha(\vec k,z)&=&\frac{1}{k^2\Omega_c}\int {\rm d^2}\vec q\, e^{i2\pi\frac{z}{k}\vec k\vec q}\nn
&& \times\,{\rm A_d}\lp \frac{f_e}{k}\vec q_+ -\vec b\rp {\rm A^*_d}\lp \frac{f_e}{k}\vec q_- -\vec b\rp\nn
&&\times\lb {\rm A_c}\lp-\frac{f_c}{k}\vec q_+\rp+{\rm A^*_c}\lp-\frac{f_c}{k}\vec q_-\rp\rb,
\eea
and the phase transfer function reads
\bea\label{eq:PSF_small_phase}
\tilde P_\varphi(\vec k,z)&=&-\frac{i}{k^2\Omega_c}\int {\rm d^2}\vec q\, e^{i2\pi\frac{z}{k}\vec k\vec q}\nn
&&\times\,{\rm A_d}\lp \frac{f_e}{k}\vec q_+ -\vec b\rp {\rm A^*_d}\lp \frac{f_e}{k}\vec q_- -\vec b\rp\nn
&&\times\lb {\rm A_c}\lp-\frac{f_c}{k}\vec q_+\rp-{\rm A^*_c}\lp-\frac{f_c}{k}\vec q_-\rp\rb.
\eea
These functions are calculated in Appendix in the case when both illumination
and detection apertures are square in shape, and the numerical aperture of
illumination does not exceed that of detection, ${\rm NA_i}\le {\rm NA_d}$. Similar analysis of the transfer functions describing other phase-sensitive optical systems has been presented in Refs.~\cite{Rose1977,Hamilton1984,Streibl1985}.

\begin{figure}[htbp]
	\centering
    \includegraphics[width=8cm]{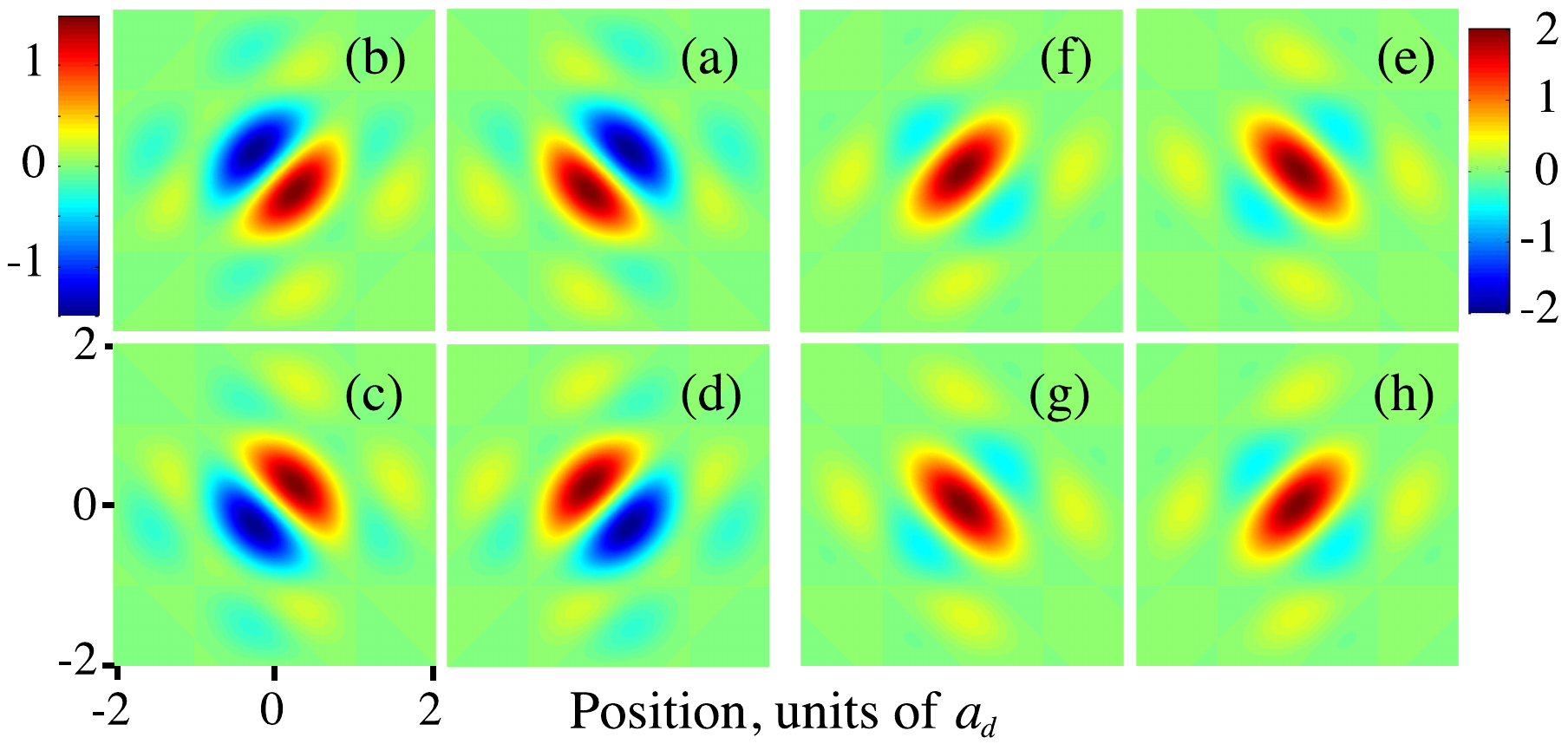} 
     \caption{(a)-(d) Spatial dependence of the phase spread functions $a_i^2 P_\varphi(\vec r,z=0)$ in quadrants 1-4, respectively; (e)-(h) the intensity spread functions $a_i^2 P_\alpha(\vec r,z=0)$ in quadrants 1-4, respectively. The functions are calculated for the in-focus position $z=0$ and detection aperture ${\rm NA_d}=2{\rm NA_i}$. Lateral positions are in units of $a_d$.}
\label{fig:PhaseIntensitySFzero}
\end{figure}

Illustrations of the spread functions of phase and intensity are shown in
Fig.~\ref{fig:PhaseIntensitySFzero} for the specific case ${\rm NA_d}=2{\rm
NA_i}$, and for the in-focus position $z=0$ (see Appendix for details). We
observe that the spatial resolution associated with the transfer functions is
$a_d=\lambda/(2{\rm NA_d})$, determined by the detection numerical aperture
${\rm NA_d}$ (also see below a detailed discussion of the transfer functions).
The phase spread function vanishes when averaged over the detection plane,
indicating that it is sensitive to phase variations only. The average of the
intensity spread function is a constant $\int d^2\vec {\rm r}\, P_\alpha(\vec
r,z)=1/2$ independent of axial position $z$.

Eqs.~(\ref{eq:PSF_0}) and~(\ref{eq:PSF_small_phase}) are some of the main
results of this work. Though we have taken as examples square apertures here,
they are valid for arbitrary shapes of the detection and illumination
apertures and for arbitrary defocus positions $z$.

\subsection{Phase-gradient expansion}

In the limit of small wavevectors $\vec k\to 0$ we recover the approximation
of smooth phase gradients~(\ref{eq:intensity_image_smooth}). Indeed, since our
optical system can only detect variations of phase, the zero frequency term of
Eq.~(\ref{eq:PSF_small_phase}), corresponding to a constant phase, vanishes,
$\tilde P_\varphi(\vec k=0,z)=0$. The first non-vanishing term is obtained by
Taylor expanding~(\ref{eq:PSF_0}) and~(\ref{eq:PSF_small_phase}) at small
spatial frequencies, leading to
\bea\label{eq:intensity_image_smooth_small}
I_d(\vec r)&\approx&\frac{I_0}{f_c^2\Omega_c}\int {\rm d^2}\vec r' \left |{\rm A_d}\lp\frac{f_e}{f_c}\vec r'-\vec b\rp\right|^2\nn
&&\times\lb 1-2\alpha(\vec r)-\frac{f_c}{2\pi k}\nabla_{\vec r}\varphi(\vec r)\nabla_{\vec r'}\rb{\rm A_c}\lp -\vec r'\rp,
\eea
which, in fact, is the Taylor expansion of
Eq.~(\ref{eq:intensity_image_smooth}) in the limit of small phase gradients
(see details in Appendix). Specifically, when a square illumination aperture
is smaller than the detection aperture of each lens of a quatrefoil
assembly~(\ref{eq:quatre-foil lens}), we obtain the small phase-gradient
approximation of Eq.~(\ref{eq:int_smooth_grad})
\be\label{eq:int_smooth_small_grad}
I_d(\vec r)\approx\frac{I_0}{4}\lb 1-2\alpha(\vec r)\pm\frac{a_{i}}{\pi}\p_x\varphi(\vec r)\pm\frac{a_{i}}{\pi}\p_y\varphi(\vec r)\rb.
\ee
The phase gradients are orthogonal here to the edges of illumination aperture.
As before, the signs of phase gradients are defined by the detection lens
quadrant. Phase gradients are obtained using the linear combinations of
intensities~(\ref{eq:tilts}).

\section{In-focus phase imaging}
The range of spatial frequencies where Eq.~(\ref{eq:reconstruct_phase_smooth})
remains valid can be estimated by studying the structure of the phase transfer
function~(\ref{eq:PSF_small_phase}) beyond the Taylor expansion. We assume
that a phase sample is located at the focal plane $z=0$ and consider a square
illumination aperture that is smaller than the detection aperture of each lens
in a quatrefoil assembly~(\ref{eq:quatre-foil lens}). As an example, we assume
that the detection aperture is a square of size $2d_p\times 2d_p$. The
corresponding numerical aperture is given by ${\rm NA_d}=d_p/f_e$.

Having derived the transfer functions~(\ref{eq:PSF_0})
and~(\ref{eq:PSF_small_phase}), we can now refine our phase imaging
method~(\ref{eq:reconstruct_phase_smooth}). Previously we made use of
Eq.~(\ref{eq:phase_gradients}) to relate local light tilts to sample phase
variations. Here, we formally define light tilts according to
Eq.~(\ref{eq:tilts}) and derive the relations between the tilts and phase
\bea
\theta_x(\vec  r)&=&{\rm NA_i}\int {\rm d^2}\vec r_0\, P^x_\theta(\vec r-\vec r_0)\varphi(\vec r_0),\nn
\theta_y(\vec  r)&=&{\rm NA_i}\int {\rm d^2}\vec r_0\, P^y_\theta(\vec r-\vec r_0)\varphi(\vec r_0).
\eea
The spread functions of the tilts are linear combinations of the spread
functions of intensities $P^{(m)}_\varphi(\vec r)$, where $m=1..4$ identifies
the quadrant of the lens assembly:
\bea
P^x_\theta&=&P^{(1)}_\varphi+P^{(4)}_\varphi- P^{(2)}_\varphi-P^{(3)}_\varphi,\nn
P^y_\theta&=&P^{(1)}_\varphi+P^{(2)}_\varphi- P^{(3)}_\varphi-P^{(4)}_\varphi.
\eea

The transfer functions of the tilts $\tilde P^x_{\theta}(\vec k)$ and $\tilde
P^y_{\theta}(\vec k)$ at the in-focus position $z=0$ are obtained by
substituting Eq.~(\ref{eq:intensity_total}) in~(\ref{eq:tilts}) and using the
Fourier transform:
\be\label{eq:tilts_fourier}
\tilde\theta_x(\vec k)={\rm NA_i}\tilde P^x_\theta(\vec k)\tilde\varphi(\vec k),\quad\tilde\theta_y(\vec k)={\rm NA_i}\tilde P^y_\theta(\vec k)\tilde\varphi(\vec k),
\ee
We substitute Eq.~(\ref{eq:PSF_small_phase}) into Fourier-transformed
Eqs.~(\ref{eq:tilts}) and derive separable representations
\bea\label{eq:SFTilts}
\tilde P^x_\theta(\vec k)=2i\, \tilde g_-(k_x)\tilde g_+(k_y),\quad
\tilde P^y_\theta(\vec k)=2i\, \tilde g_-(k_y)\tilde g_+(k_x),
\eea
where the functions $\tilde g_-(q)$ and $\tilde g_+(q)$ are illustrated in
Fig.~\ref{fig:TiltSF} and explicitly expressed in Appendix.

\begin{figure}[htbp]
	\centering
    \includegraphics[width=8cm]{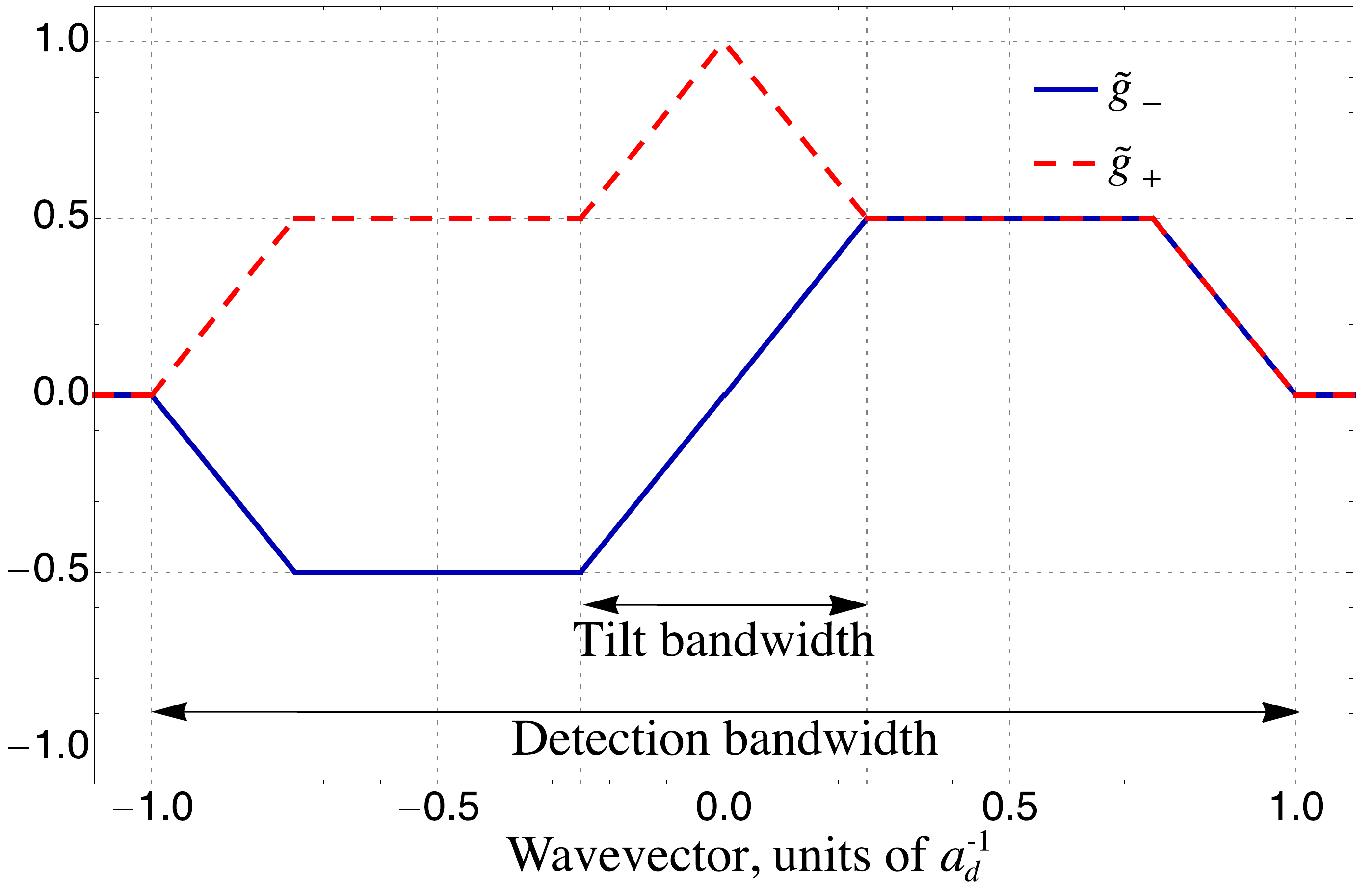}
     \caption{Plots of $\tilde g_-$ and $\tilde g_+$ for in-focus imaging when the detection aperture is a square and ${\rm NA_d}=2{\rm NA_i}$.}
\label{fig:TiltSF}
\end{figure}

We can understand the behavior of these functions qualitatively, again from
Fig.~\ref{fig:setup}(b). Specifically, the function $\tilde g_-(q)$ is
proportional to the tilt angle along one direction. Let us consider a
prism-like sample that refracts light along $y$-axis. A positive tilt is
imposed by the prism $\theta_y=k_y/k\ge 0$ along $y$-axis. When the tilt is
relatively small, the image shifts upwards, and the two upper lenses of the
quatrefoil assembly collect more light than the lower ones. Correspondingly,
function $\tilde g_-(k_y)$ grows linearly with $k_y$ until the tilt angle
reaches the limiting value of the tilt range, $\theta_y= {\rm NA_i}$. For
larger angles, ${\rm NA_i}<\theta_y\le {\rm NA_d}-{\rm NA_i}$, the lower
lenses collect no light, while the upper ones collect the maximum power. The
power does not change until the image of the illumination aperture reaches the
top edge of the lens assembly at $\theta_y={\rm NA_d}-{\rm NA_i}$. Still at
larger angles of refraction, the intensity collected by the upper lenses
begins to decrease until it falls to zero when the image of illumination
aperture leaves the lens assembly altogether at $\theta_y={\rm NA_d}$. Similar
arguments apply to negative tilts $\theta_y\le 0$, and it is straightforward
to see that $\tilde g_-(k_y)$ is an odd function of $k_y$. The optical
bandwidth of the system is defined here by the detection aperture,
$|k_{x,y}|\le 1/a_d$, where $a_d=\lambda/(2{\rm NA_d})$. The function $\tilde
g_-$ is shown in Fig.~\ref{fig:TiltSF}.

\begin{figure}[htbp]
	\centering
    \includegraphics[width=8cm]{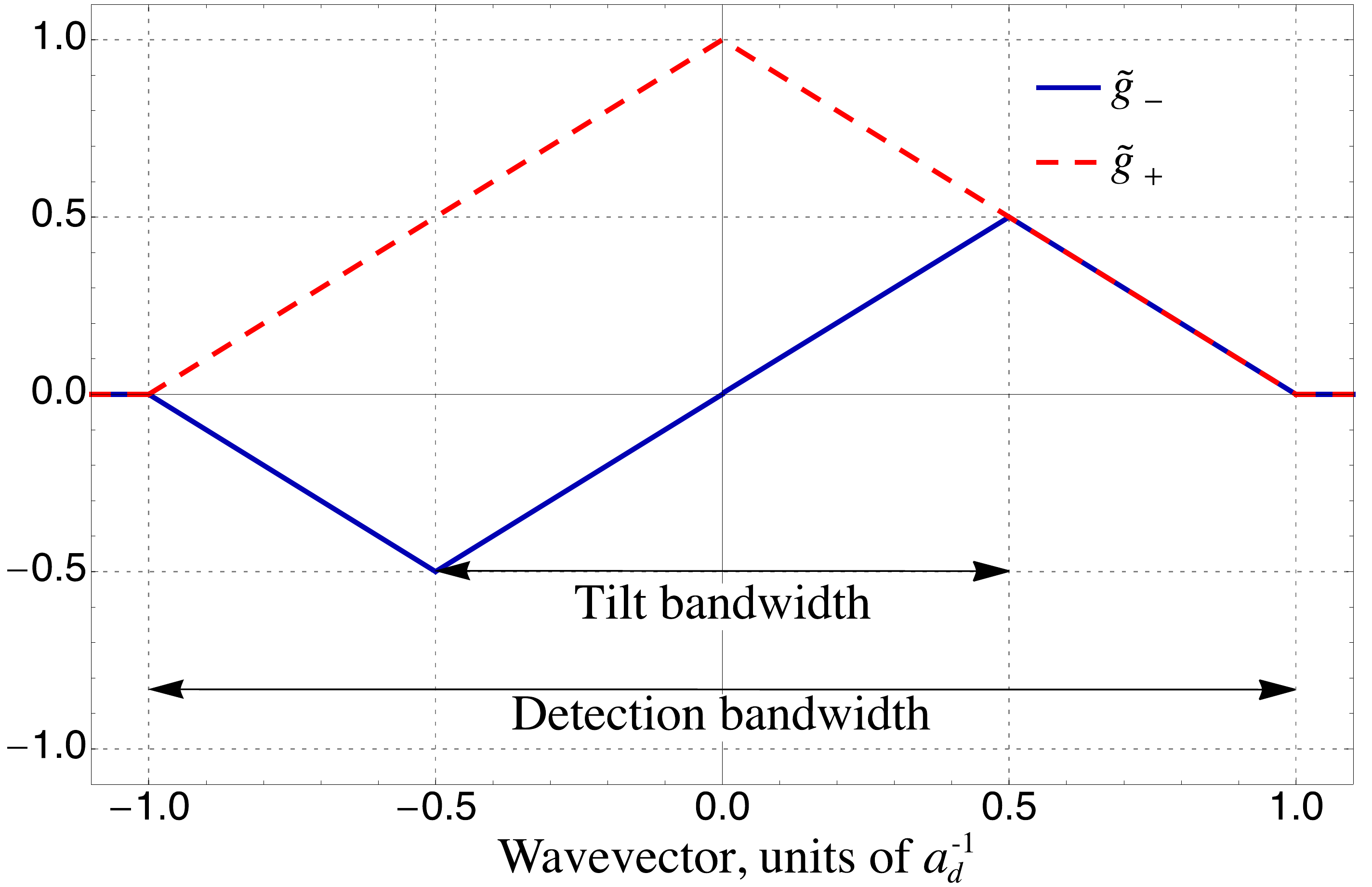}
     \caption{Plots of $\tilde g_-$ and $\tilde g_+$ for in-focus imaging when the detection aperture is a square and ${\rm NA_d}={\rm NA_i}$.}
\label{fig:TiltSFNAdEqNAi}
\end{figure}

The detection and illumination apertures completely characterize the
light-tilt transfer functions. As we discuss below, to achieve
detection-limited resolution for imaging out-of-focus phase objects, the
numerical apertures of illumination and detection should be matched, ${\rm
NA_i}={\rm NA_d}$. Fig.~\ref{fig:TiltSFNAdEqNAi} illustrates the transfer
functions in this case. In contrast to Fig.~\ref{fig:TiltSF}, the range of
frequencies for which $\tilde g_-=1/2$ has been reduced to a point.

\begin{figure}[htbp]
	\centering
    \includegraphics[width=8cm]{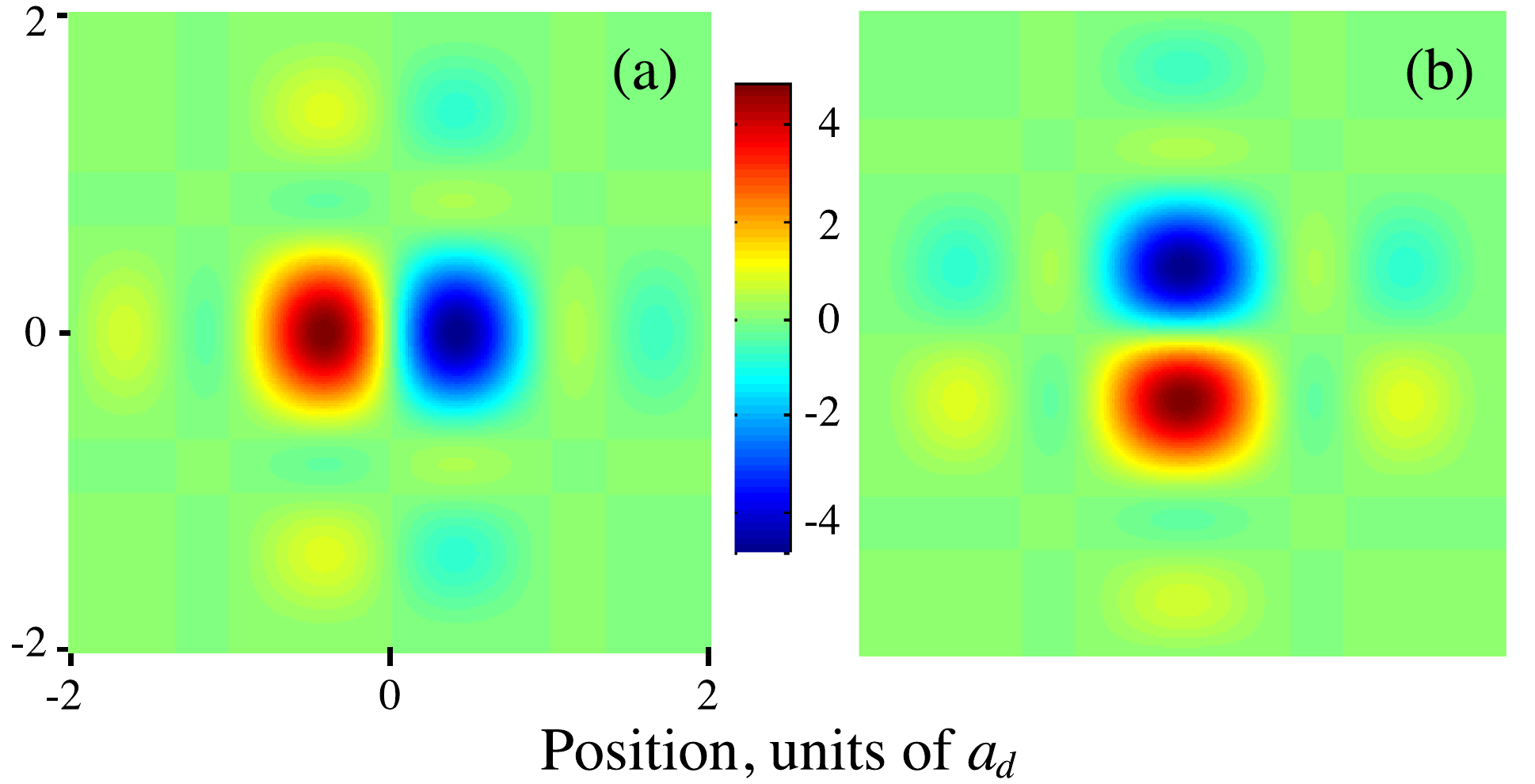}
     \caption{Spatial dependence of the tilt spread functions: $a_i^2P^x_\theta(\vec r,z=0)$ (a), and $a_i^2P^y_\theta(\vec r,z=0)$ (b), calculated for the in-focus position $z=0$, when the detection aperture is a square and ${\rm NA_d}=2{\rm NA_i}$. Lateral positions are in units of $a_d$.}
\label{fig:TiltsSpreadSpace}
\end{figure}

\subsection{Spatial frequencies limited by the illumination aperture}

To begin, we consider spatial frequencies within the tilt range of our
system, namely $|k_{x,y}|\le 1/(2a_i)$, as defined by the illumination
numerical aperture. The functions $\tilde g_-(k_{x,y})=k_{x,y} a_i$ and
$\tilde g_+(k_{x,y})=1-|k_{x,y} a_i|$ are then independent of the detection
aperture. The corresponding Fourier components of the tilts~(\ref{eq:SFTilts})
are given by
\bea\label{eq:tilts_in_focus}
\tilde\theta_x(\vec k)&=&i\frac{k_x}{k} (1-|k_y a_i|)\tilde\varphi(\vec k),\nn \tilde\theta_y(\vec k)&=&i\frac{k_y}{k} (1-|k_x a_i|)\tilde\varphi(\vec k),
\eea
where $|k_{x,y}|\le 1/(2a_i)$. These equations uniquely map the directions of
light rays $k_{x,y}/k$ to the tilt angles along the same axes in proportion to
the Fourier components of the phase. These may be compared with
Eqs.~(\ref{eq:phase_gradients}) utilized for our smooth phase approximation,
and are found to be equivalent except for an additional attenuation of spatial
frequencies orthogonal to the tilt directions.

Phase is extracted from~(\ref{eq:tilts_in_focus}) using a linear combination
of the tilts
\be\label{eq:fourier_phase_in_focus}
\tilde\varphi(\vec k) = -ik\,\frac{\tilde\theta_x(\vec k)+i\tilde\theta_y(\vec k)}{k_x (1-|k_y a_i|)+ik_y (1-|k_x a_i|)},
\ee
where $|k_{x,y}|\le 1/(2a_i)$, or, correspondingly,
\be\label{eq:phase_reconstruction}
\varphi(\vec r)=-ik\int {\rm d^2} \vec k\,e^{i2\pi \vec k\vec r}\, \frac{\tilde\theta_x(\vec k)+i\tilde\theta_y(\vec k)}{k_x (1-|k_y a_i|)+ik_y (1-|k_x a_i|)}, 
\ee
where integration extends over the tilt range $|k_{x,y}|\le 1/(2a_i)$.
In this derivation we assumed that the spatial support of $\varphi(\vec r)$ is
finite within the imaging field of view. We note that
Eq.~(\ref{eq:phase_reconstruction}) is applicable to arbitrary shapes of the
detection lens provided the associated aperture exceeds the numerical aperture
of illumination. The phase reconstruction~(\ref{eq:phase_reconstruction})
should be used when the illumination aperture is well characterized, while the
detection aperture is not precisely known. However, the method does not
account for the spatial frequencies beyond the tilt range.

\subsection{Spatial frequencies limited by the detection aperture}

The resolution of phase reconstruction can be improved if the detection
aperture is precisely known. We use the known spread functions to deconvolve
light tilts and obtain the phase
\be\label{eq:phase_reconstruction_precise}
\varphi(\vec r)=\frac{1}{{\rm NA_i}}\int {\rm d^2} \vec k\,e^{i2\pi \vec k\vec r}\, \frac{\tilde\theta_x(\vec k)+i\tilde\theta_y(\vec k)}{\tilde P^x_\theta(\vec k)+i\tilde P^y_\theta(\vec k)}, 
\ee  
where integration now extends over the detection bandwidth. For a square
detection aperture, the spread functions of light tilts are given in
Appendix. Corresponding illustrations of the tilt spread functions
$P^x_\theta(\vec r)$ and $P^y_\theta(\vec r)$ are shown in
Fig.~\ref{fig:TiltsSpreadSpace} for the in-focus position $z=0$. As can be
observed, $P^x_\theta(\vec r)$ is even with respect to $y$-axis and odd with
respect to $x$-axis, and for the function $P^y_\theta(\vec r)$ vice-versa.

\begin{figure}[htbp]
	\centering
    \includegraphics[width=8cm]{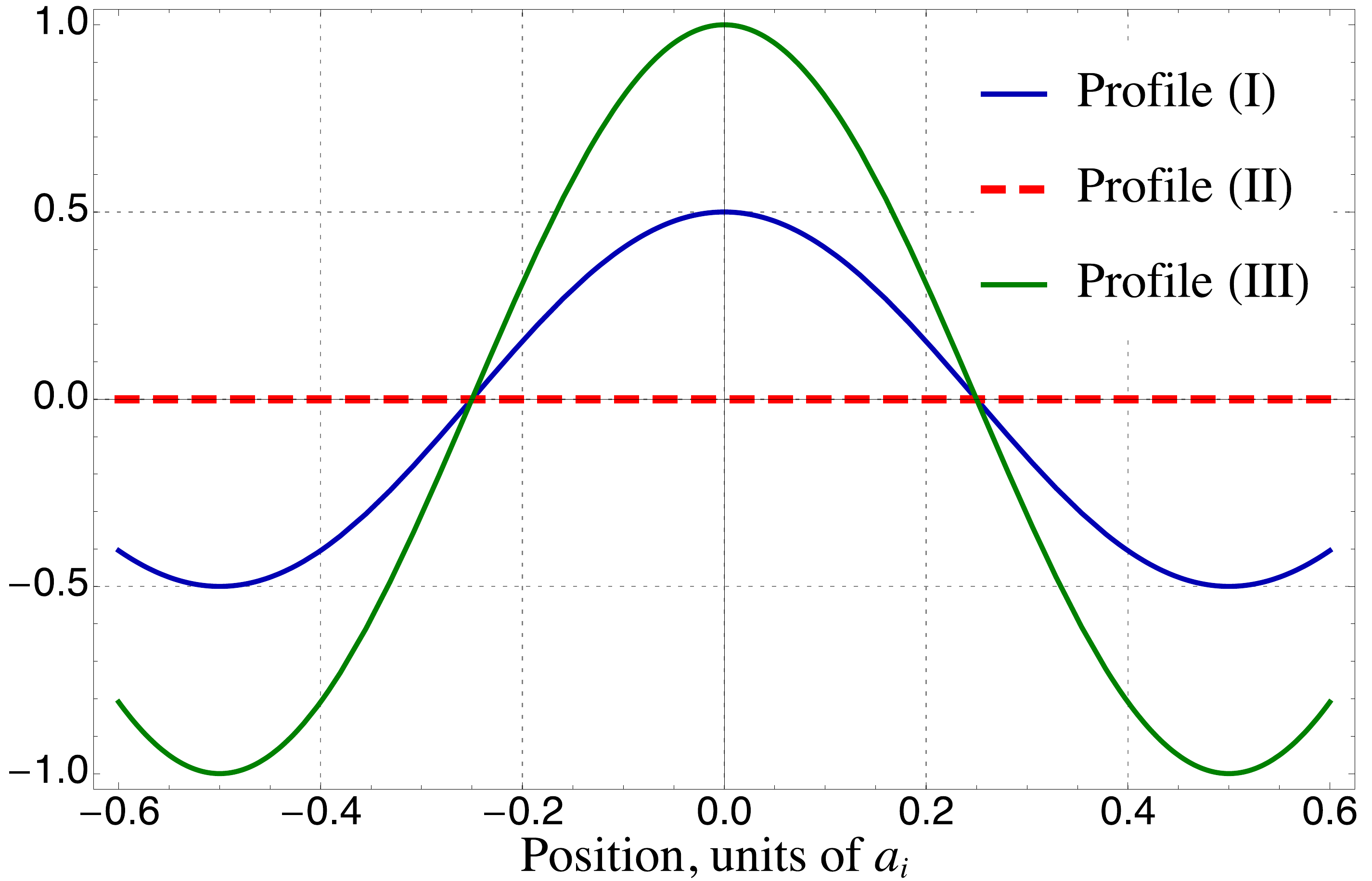}
     \caption{Reconstruction of a periodic phase profile $\varphi(x)/\varphi_0$ in Eq.~(\ref{eq:HarmonicProfile}) when $Q=1/a_i$ and ${\rm NA_d}=2{\rm NA_i}$ using different methods: Eq.~(\ref{eq:reconstruct_phase_smooth}) (solid blue line), Eq.~(\ref{eq:phase_reconstruction}) (red dashed line), and Eq.~(\ref{eq:phase_reconstruction_precise}) (green solid line). The exact profile is depicted by the green solid line. Lateral position is in units of $a_i$.}
\label{fig:PhaseRecsHarmonics}
\end{figure}

Let us compare the reconstruction methods~(\ref{eq:reconstruct_phase_smooth}),
(\ref{eq:phase_reconstruction}) and (\ref{eq:phase_reconstruction_precise}) in
the simple case of a one-dimensional periodic phase profile
\be\label{eq:HarmonicProfile}
\varphi(x)=\varphi_0\cos(2\pi Q x),
\ee
where $Q\ge 0$ is the periodic spatial frequency. It is easy to verify that
when $Q\le 1/(2a_i)$ different phase reconstruction methods yield an identical
profile $\varphi_{rec}(x)=\varphi(x)$. However, for larger frequencies the
reconstructions are different. We consider for example frequencies within the
range $1/(2a_i)\le Q\le 1/a_d-1/a_i$. The tilt transfer function is then
constant $\tilde g_{-}(\pm Q)=\pm 1/2$ (see Fig.~\ref{fig:TiltSF}). The
reconstruction results are summarized in Fig.~\ref{fig:PhaseRecsHarmonics},
where ${\rm NA_d}=2{\rm NA_i}$ and frequency $Q=1/a_i$. According to
Eq.~(\ref{eq:reconstruct_phase_smooth}) we obtain
$\varphi_{rec}(x)=\psi\cos(2\pi Qx)$, where the reconstructed amplitude
$\psi=\varphi_0/(2a_i Q)$ incorrectly depends on the spatial frequency $Q$
instead of being the constant $\varphi_0$ (Profile (I) in
Fig.~\ref{fig:PhaseRecsHarmonics}). This inconsistency is due to the incorrect
mapping of phase to tilts outside of the tilt range. A refined
approach~(\ref{eq:phase_reconstruction}) is only sensitive to spatial
frequencies within the tilt range, so when $Q>1/(2a_i)$, the reconstructed
profile $\varphi_{rec}(x)=0$ (Profile (II) in
Fig.~\ref{fig:PhaseRecsHarmonics}) is also incorrect. Finally,
Eq.~(\ref{eq:phase_reconstruction_precise}) properly accounts for all spatial
frequencies within the detection bandwidth $Q\le 1/a_d$, and, as a result,
provides the correct reconstruction of the original profile
$\varphi_{rec}(x)=\varphi(x)$ (Profile (III) in
Fig.~\ref{fig:PhaseRecsHarmonics}). Thus all the spatial frequency components
of an arbitrary phase profile within the detection bandwidth can, in
principle, be found using Eq.~(\ref{eq:phase_reconstruction_precise}).
Approximate reconstructions of phase samples characterized by even broader
spatial frequency distributions are also possible, as we discuss below.

The phase reconstruction methods~(\ref{eq:phase_reconstruction})
and~(\ref{eq:phase_reconstruction_precise}) are the main results of this work.
Experiments reported in Refs.~\cite{Mertz2012,Barankov2013} employed the
simplified method~(\ref{eq:reconstruct_phase_smooth}) of phase reconstruction.
Since the spatial frequencies of the imaged samples were safely confined to
the tilt range, the reconstruction~(\ref{eq:reconstruct_phase_smooth})
provided good approximations of phase profiles.

\subsection{Imaging a phase step}
It is instructive to apply the above formalism to the imaging of a phase,
which manifestly is not smooth and features high spatial frequencies. The
strong light diffraction from the discontinuity allows to visualize spatial
resolution of our phase imaging method. For simplicity we consider a
one-dimensional step
\be\label{eq:phase_step}
\varphi(\vec r)=\varphi_0\, {\rm H}(x),
\ee
where the amplitude is small, $|\varphi_0|\ll 1$, and the Heaviside function ${\rm H}(x) =1$ at $x\ge 0$, and ${\rm H}(x)=0$ at $x<0$.

\begin{figure}[htbp]
	\centering
    \includegraphics[width=8cm]{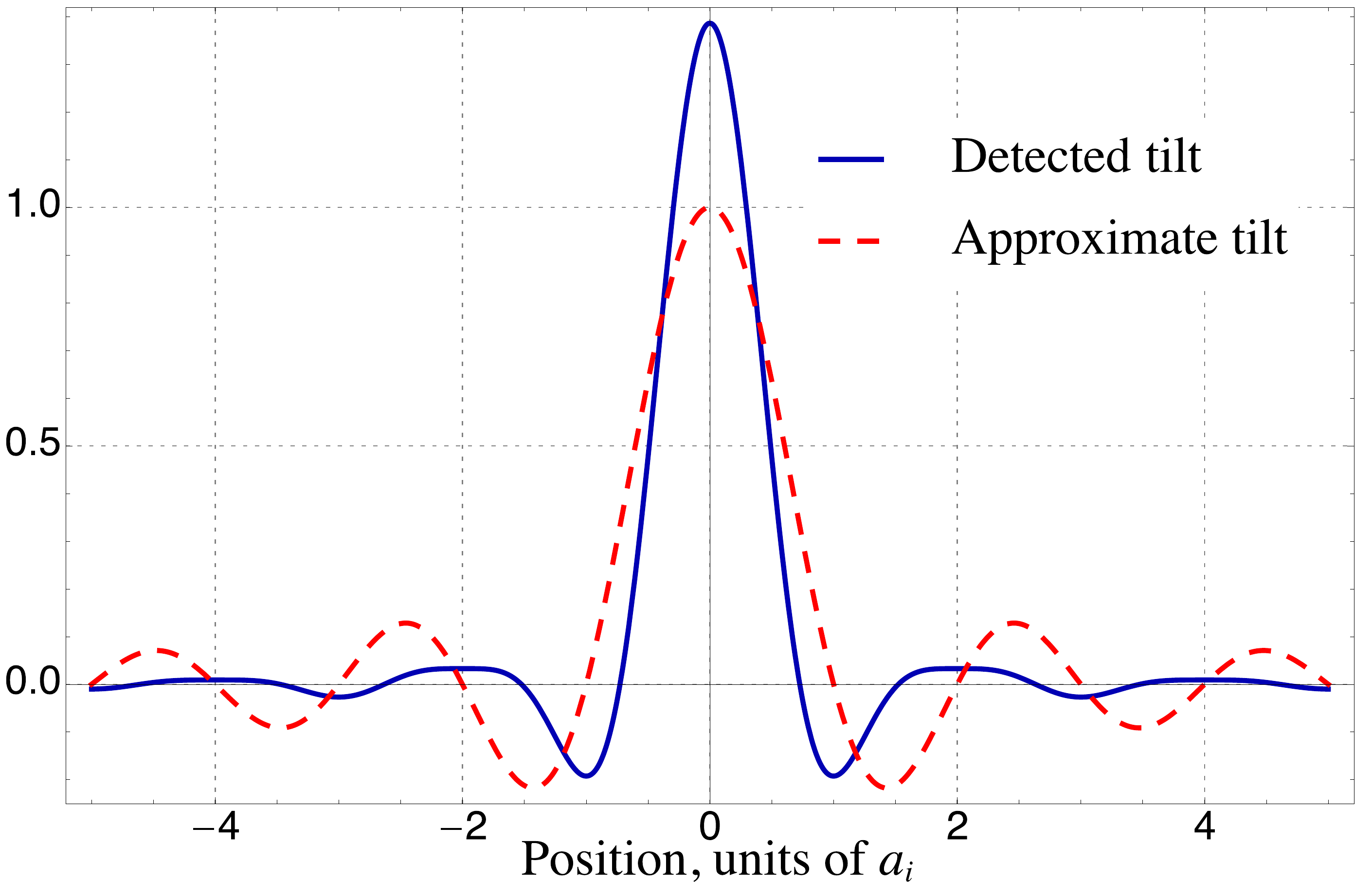}
     \caption{Detected and approximate light tilts $\theta_x/\theta_0$ as a function of position when imaging a phase step~(\ref{eq:phase_step})and in the case of matched numerical apertures ${\rm NA_d}={\rm NA_i}$. Lateral position is in units of $a_i$.}
\label{fig:theta_step}
\end{figure}

The Fourier components $\tilde\varphi(\vec k)$ are calculated using standard
regularization, in which we add an infinitesimal imaginary part to the
wavevector $k_x\to k_x-i0$
\be\label{eq:phase_step_fourier}
\tilde\varphi(\vec k)=\frac{\varphi_0}{2\pi i (k_x-i0)}\delta(k_y).
\ee 
The Fourier components of the detected light tilts~(\ref{eq:tilts}) resulting
from the object are limited by the detection aperture
\be\label{eq:step_tilt_detected_fourier}
\tilde\theta_x(\vec k)=\theta_0\frac{\tilde g_-(k_x)}{k_x}\delta(k_y),\quad \tilde\theta_y(\vec k)=0,\quad |k_x|\le 1/a_d,
\ee 
where we define the light tilt $\theta_0=(\varphi_0/\pi){\rm NA_i}$. The
discontinuity~(\ref{eq:phase_step_fourier}) diffracts light in all directions
along $x$-axis, and the detection aperture collects only part of this light.

\begin{figure}[htbp]
	\centering
    \includegraphics[width=8cm]{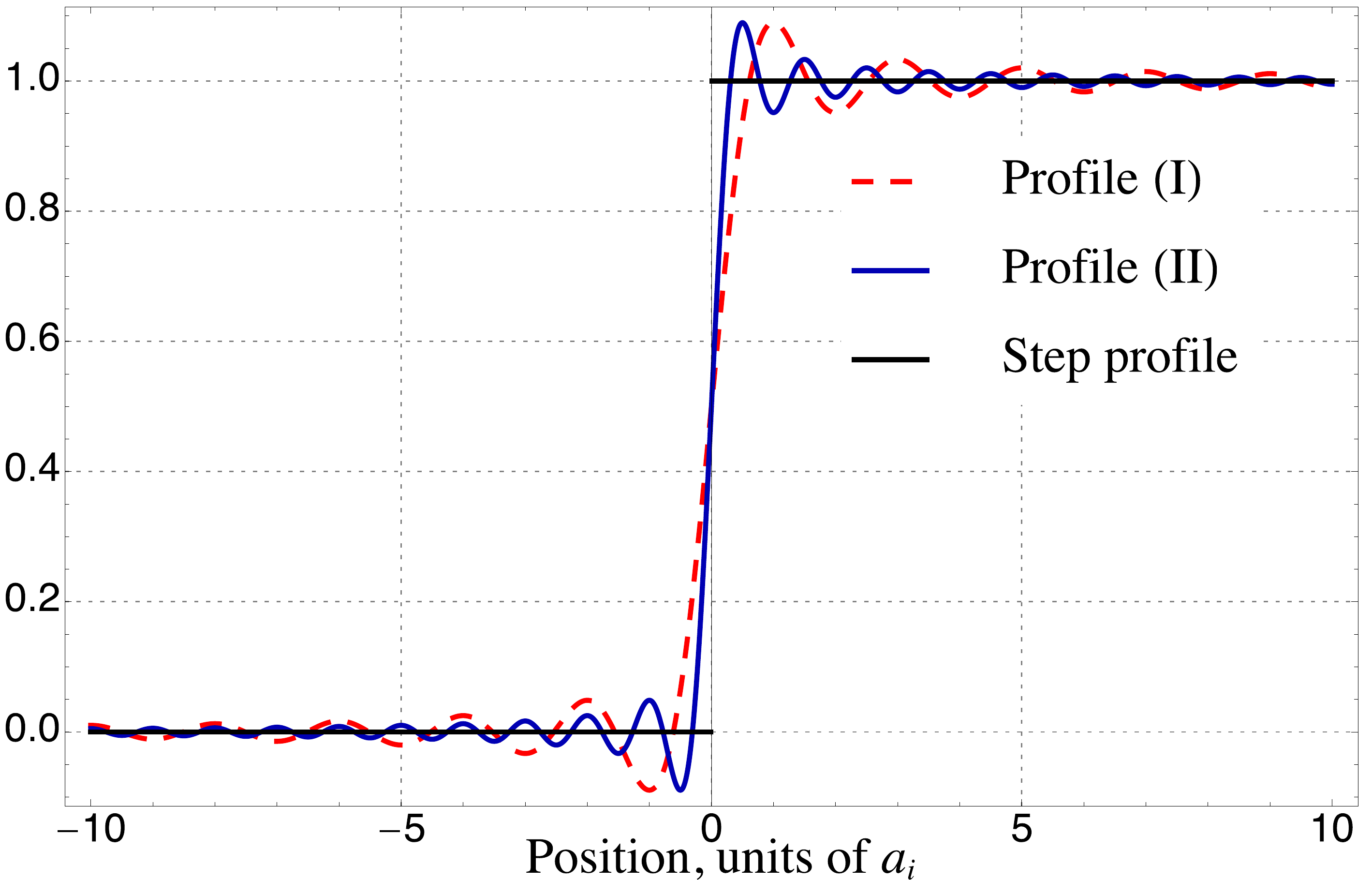}
     \caption{Step profiles $\varphi(x)/\varphi_0$ reconstructed using Eq.~(\ref{eq:reconstruct_phase_smooth}) (Profile I, red dashed line), and Eq.~(\ref{eq:phase_reconstruction}) (Profile II, solid blue line), with detection aperture ${\rm NA_d}={\rm NA_i}$. The profiles approximate a step function~(\ref{eq:phase_step}) shown as in solid black.  Lateral position is in units of $a_d$.}
\label{fig:phase_step}
\end{figure}

In real space the detected tilt angle becomes
\be\label{eq:step_tilt_detected}
\theta_x(x)=\theta_0\int_{-1/a_d}^{1/a_d} {\rm d}k_x\, e^{i2\pi k_x x} \tilde g_-(k_x)/k_x, \quad \theta_y(\vec r)=0.
\ee
The dependence of the detected light tilt on the position along $x$-axis is
plotted as a solid blue line in Fig.~\ref{fig:theta_step}, assuming a square
detection aperture of numerical aperture ${\rm NA_d}={\rm NA_i}$. For
comparison we also plot an approximate tilt (dashed red line) given by
$\theta_x(x)=\theta_0\,{\rm sinc}(\pi x/a_i)$, which is obtained from the
detected tilt~(\ref{eq:step_tilt_detected_fourier}) by setting to zero the
spatial frequencies $|k_x|\ge 1/(2a_i)$. We interpret the peak of the
approximate tilt as follows. As a light ray refracts from the step, it
effectively traverses an optical path $h=\varphi_0\lambda/(2\pi)$ within a
lateral spot of size estimated by the resolution scale $a_i=\lambda/(2{\rm
NA_i})$. The effective angle of refraction (i.e. tilt) is then estimated as
the ratio of the optical path to the spot size, $\theta=h/a_i$, which
coincides with the peak value $\theta_0=(\varphi_0/\pi){\rm NA_i}$ of the
approximate tilt. The detected tilt (shown in blue solid line) is obtained
similarly using the detection-limited resolution. Therefore, the peak detected
tilt is expected to be proportionally larger. However, since the transfer
function $\tilde g_-$ in Fig.~(\ref{fig:TiltSFNAdEqNAi}) is not linear, the
peak value of the detected tilt is only somewhat larger than the peak of the
approximate tilt.

Results of phase reconstruction are summarized in~Fig.~\ref{fig:phase_step}.
All reconstruction methods define the phase profiles up to an arbitrary
constant, which we choose to be $\varphi(0)=\varphi_0/2$. The low-frequency
part of the tilt $\tilde\theta_x(\vec k)=\theta_0 a_i\delta(k_y)$ valid within
the tilt range $|k_x|\le 1/(2a_i)$, is employed in the
reconstruction~(\ref{eq:phase_reconstruction}). The corresponding profile (I)
is $\varphi(x)/\varphi_0=1/2+(1/\pi) {\rm Si}(\pi x/a_i)$, where ${\rm
Si}(z)=\int_0^z {\rm d}q\sin (q)/q$, shown as a dashed red line. The
oscillations near the discontinuity $x=0$, known as the Gibbs phenomenon, are
attributed to the finite range $|k_x|\le 1/(2a_i)$ of the integration
in~(\ref{eq:phase_reconstruction}). The
reconstruction~(\ref{eq:phase_reconstruction_precise}) is given by
$\varphi(x)/\varphi_0=1/2+(1/\pi){\rm Si}(2\pi x/a_i)$, and plotted as the
solid blue line Profile (II). We note that in the considered case $a_i=a_d$.
The oscillations around the step identify the spatial resolution of our system
since they are a result of the finite detection bandwidth. Naturally, as the
detection bandwidth is increased, the profile approaches the step function.

\section{Out-of-focus phase imaging}\label{Section:SmallGradientsOutOfFocus}

In this section we analyze phase imaging of out-of-focus objects located a
distance $z$ from the focal plane. The acquired images in this case exhibit
both blurring and quadrant-dependent lateral shearing compared to in-focus
imaging. As in light field imaging~\cite{Levoy2006}, we make use of the fact
that shearing provides information on the defocus position and can be
compensated for. The blurring of images for large defocus, however, cannot be
compensated and degrades spatial resolution. We show that out-of-focus phase
profiles can nevertheless, in principle, be reconstructed, though with a
reduced spatial resolution depending on the extent of defocus.

The exact relation between the numerical apertures of illumination and
detection is not critical for phase reconstruction of in-focus objects,
provided the detection aperture is accurately known. For defocused objects the
situation is different, since the two apertures both control the effects of
blurring and shearing. The case of matched numerical apertures ${\rm
NA_i}={\rm NA_d}$ is special, since in that case lateral shearing observed in
different quadrants at the optical resolution uniquely identifies the defocus
distance (see below and also Appendix). In the unbalanced case ${\rm
NA_i}<{\rm NA_d}$, the situation is more complicated. The problem comes from
the spectral components of intensities outside the tilt bandwidth, since they
shear differently than the spectral components within the bandwidth. That is
why, we consider only the frequencies within the detection bandwidth
$|k_{x,y}|\le 1/(2a_i)$, which can be accomplished by applying a low-pass
filter to the recorded images. The formalism presented below concentrates on
the detection-limited phase reconstruction for a system with matched numerical
apertures. However, it can also be applied in the unbalanced case after
numerically filtering the bandwidth of the recorded images to be within the
tilt bandwidth (see also Section~\ref{section:unbalanced out-of-focus imaging}
in Appendix).

\begin{figure}[htbp]
	\centering
    \includegraphics[width=8cm]{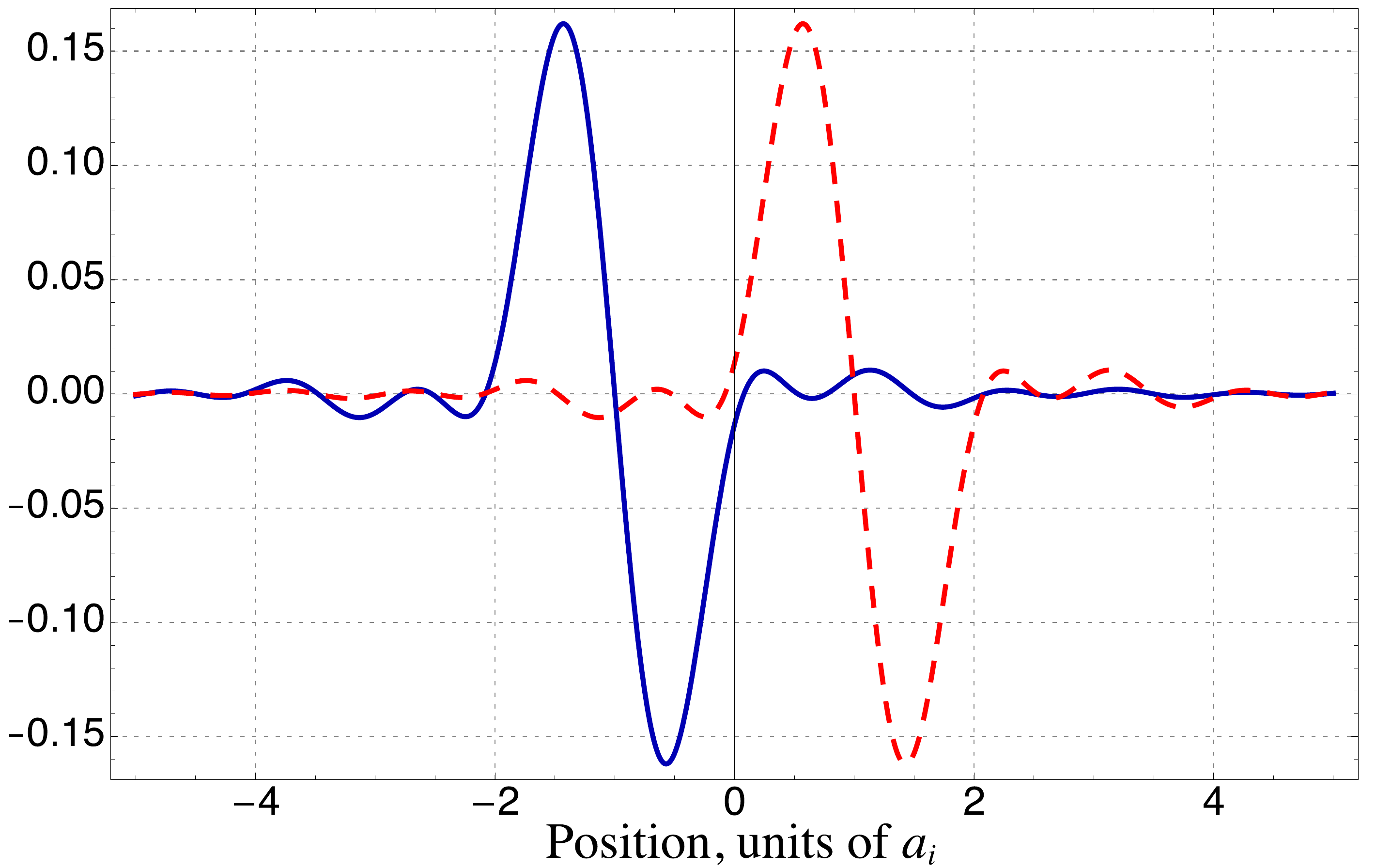}
     \caption{Averaged phase spread functions, $a_i\la P_\varphi^{(1)}(x,y)\ra_{y}$ (solid blue line) and the negative of $a_i\la P_\varphi^{(2)}(x,y)\ra_{y}$ (dashed red line), calculated in quadrants (1)  and (2), as functions of lateral position $x$ at the defocus position $z=z_i$, in the case of matched numerical apertures ${\rm NA_i}={\rm NA_d}$. The two functions are replicas of one another, shifted by $\Delta x= 2{\rm NA_i} z$. Lateral position is in units of $a_i$.}
\label{fig:PhaseSFAve12}
\end{figure}

The principle of numerical refocusing is illustrated in
Fig.~\ref{fig:PhaseSFAve12}, where we considered the case of matched numerical
apertures ${\rm NA_i}={\rm NA_d}$. We find that defocused spread functions of
quadrants (1) and (2) averaged over the $y$ coordinate, $\la
P_\varphi^{(1)}(x,y,z)\ra_{y}$ and $\la P_\varphi^{(2)}(x,y,z)\ra_{y}$, are
replicas of one another shifted along $x$-axis, up to a negative sign. Since
the intensity of an image is obtained as a
convolution~(\ref{eq:phase_sensitive_intensity}) of phase and the phase spread
function, we observe that the images of an arbitrary phase object shift
accordingly. The relative shift is proportional to the defocus distance,
$\Delta x=2{\rm NA_i}z$, and, thus, unambiguously identifies this distance.

In the two-dimensional case, to avoid ambiguities, we also need to consider
the first and fourth quadrants, and average the corresponding phase-sensitive
intensities over the $x$ coordinate. The averaged intensities depending on
coordinate $y$ are the shifted replicas of one another, up to a negative sign.
The two relations between the intensities are given by
\bea\label{eq:Intensities_sheared}
\la I^{(1)}_\varphi(x-{\rm NA_i}z,y)\ra_{y}&=&-\la I^{(2)}_\varphi(x+{\rm NA_i}z,y)\ra_{y},\nn
\la I^{(1)}_\varphi(x,y-{\rm NA_i}z)\ra_{x}&=&-\la I^{(4)}_\varphi(x,y+{\rm NA_i}z)\ra_{x},
\eea
where $\la ..\ra_{x,y}$ is the average over coordinates $x$ and $y$,
respectively, and $I^{(m)}_\varphi(\vec r)$ is a phase-sensitive part of the
intensity recorded in quadrant $m$. The intensity $I^{(m)}_\varphi(\vec r)$ is
obtained by subtracting the background from the total
intensity~(\ref{eq:intensity_total}).
Identities~(\ref{eq:Intensities_sheared}) provide, in principle, an estimate
of the defocus position $z$, for example, by maximizing the cross-correlation
function of the averaged intensities.

Light tilts adjusted for the shearing are given by
\bea\label{eq:Tilts_z}
\theta_x(\vec r,z)&=&\frac{{\rm NA_i}}{I_{tot}}\left[\begin{array}{ll} I_{1}(x_-,y_-)+I_{4}(x_-,y_+)\\
-I_{2}(x_+,y_-)-I_{3}(x_+,y_+)\end{array}\right],\nn
\theta_y(\vec r,z)&=&\frac{{\rm NA_i}}{I_{tot}}\left[\begin{array}{ll} I_{1}(x_-,y_-)+I_{2}(x_-,y_+)\\
-I_{3}(x_+,y_-)-I_{4}(x_+,y_+)\end{array}\right],
\eea
where $x_{\pm}=x\pm {\rm NA_i}z$, $y_{\pm}=y\pm {\rm NA_i}z$, $I_m(\vec r)$ is the intensity recorded in quadrant $m$, and $I_{tot}=I_{1}(x_-,y_-)+I_{2}(x_-,y_+)+I_{3}(x_+,y_-)+I_{4}(x_+,y_+)$.

\begin{figure}[htbp]
	\centering
    \includegraphics[width=8cm]{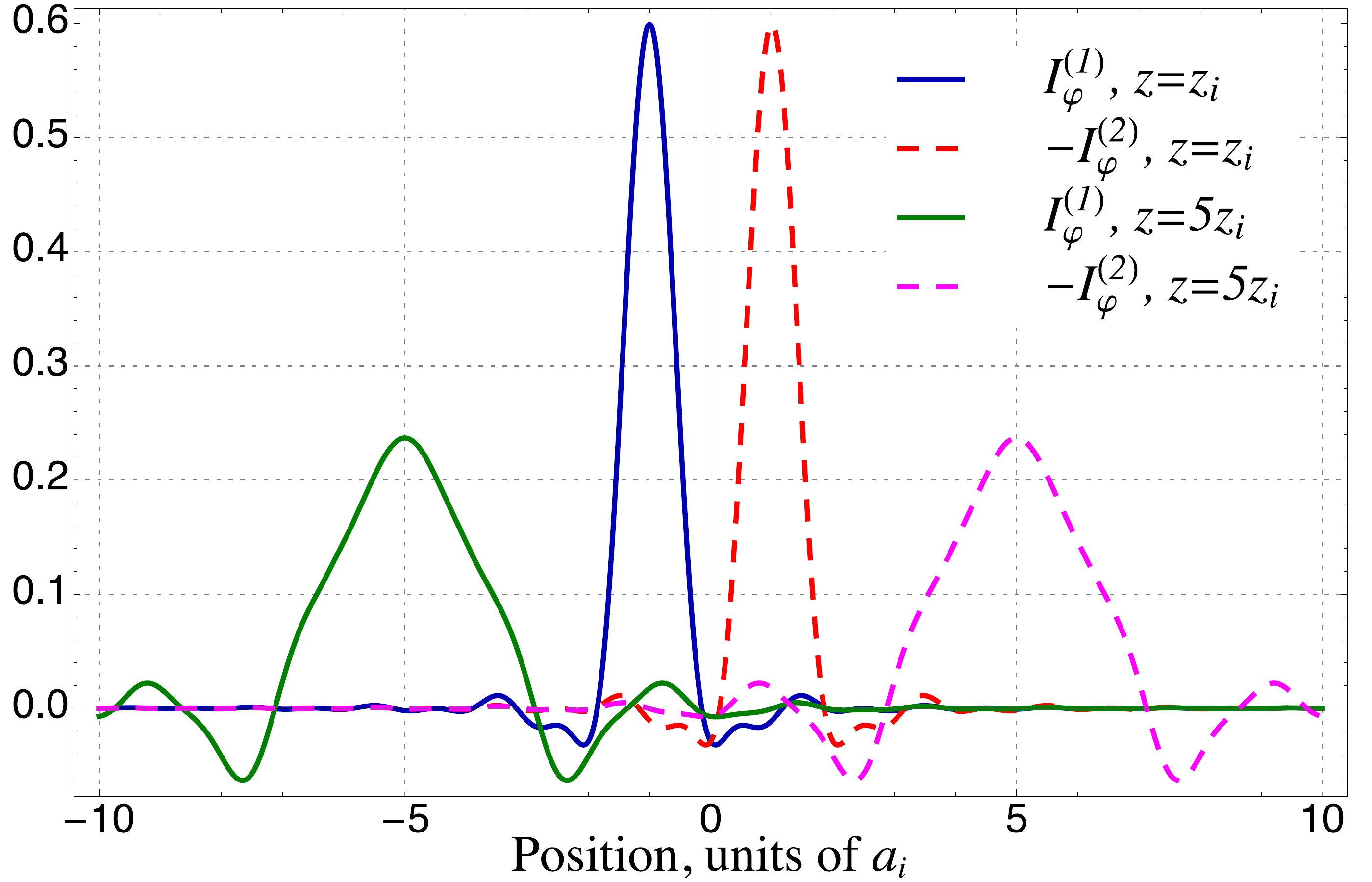}
     \caption{Intensities $I^{(1)}_\varphi(x,z)$ and negative $I^{(2)}_\varphi(x,z)$ as functions of lateral position $x$, calculated for different defocus distances and with matched numerical apertures ${\rm NA_i}={\rm NA_d}$. Solid blue line and dashed red line correspond to $z=z_i$; solid green line and dashed magenta line correspond to $z=5z_i$. Lateral position is in units of $a_i$.}
\label{fig:IntensityStepQ12}
\end{figure}

We find the Fourier components of~(\ref{eq:Tilts_z}) and use the known
transfer functions of tilts~(\ref{eq:TFTiltsMatchedApertures}) (see Appendix)
to obtain the phase of an out-of-focus object
\be\label{eq:phase_reconstruction_z}
\varphi(\vec r,z)=\frac{1}{{\rm NA_i}}\int {\rm d^2} \vec k\,e^{i2\pi \vec k\vec r}\, \frac{\tilde\theta_x(\vec k,z)+i\tilde\theta_y(\vec k,z)}{\tilde P^x_\theta(\vec k,z)+i\tilde P^y_\theta(\vec k,z)}, 
\ee  
where the integration extends over the detection bandwidth $|k_{x,y}|\le
1/a_i$, when the numerical apertures of illumination and detection match,
${\rm NA_i}={\rm NA_d}$ (i.e. $a_d=a_i$). We note that for relatively small
defocus, deconvolution is in principle possible within the detection
bandwidth, based on the transfer functions~(\ref{eq:TFTiltsMatchedApertures}).
In cases when ${\rm NA_i}<{\rm NA_d}$ the integration extends over the tilt
bandwidth $|k_{x,y}|\le 1/(2a_i)$, which leads to a reduced spatial resolution
compared to the case of matched apertures. For a relatively large defocus, the
combination of transfer functions in~(\ref{eq:phase_reconstruction_z}) has
zeros at $k_{x,y}\ne 0$. The bandwidth should be reduced to exclude these
singularities from the integration. The appearance of singularities identifies
the defocus regime where the blurring cannot be accurately compensated, which
results in reduced resolution of phase reconstruction depending on the defocus
distance.

\subsection{Imaging a defocused phase step}

\begin{figure}[htbp]
	\centering
    \includegraphics[width=8cm]{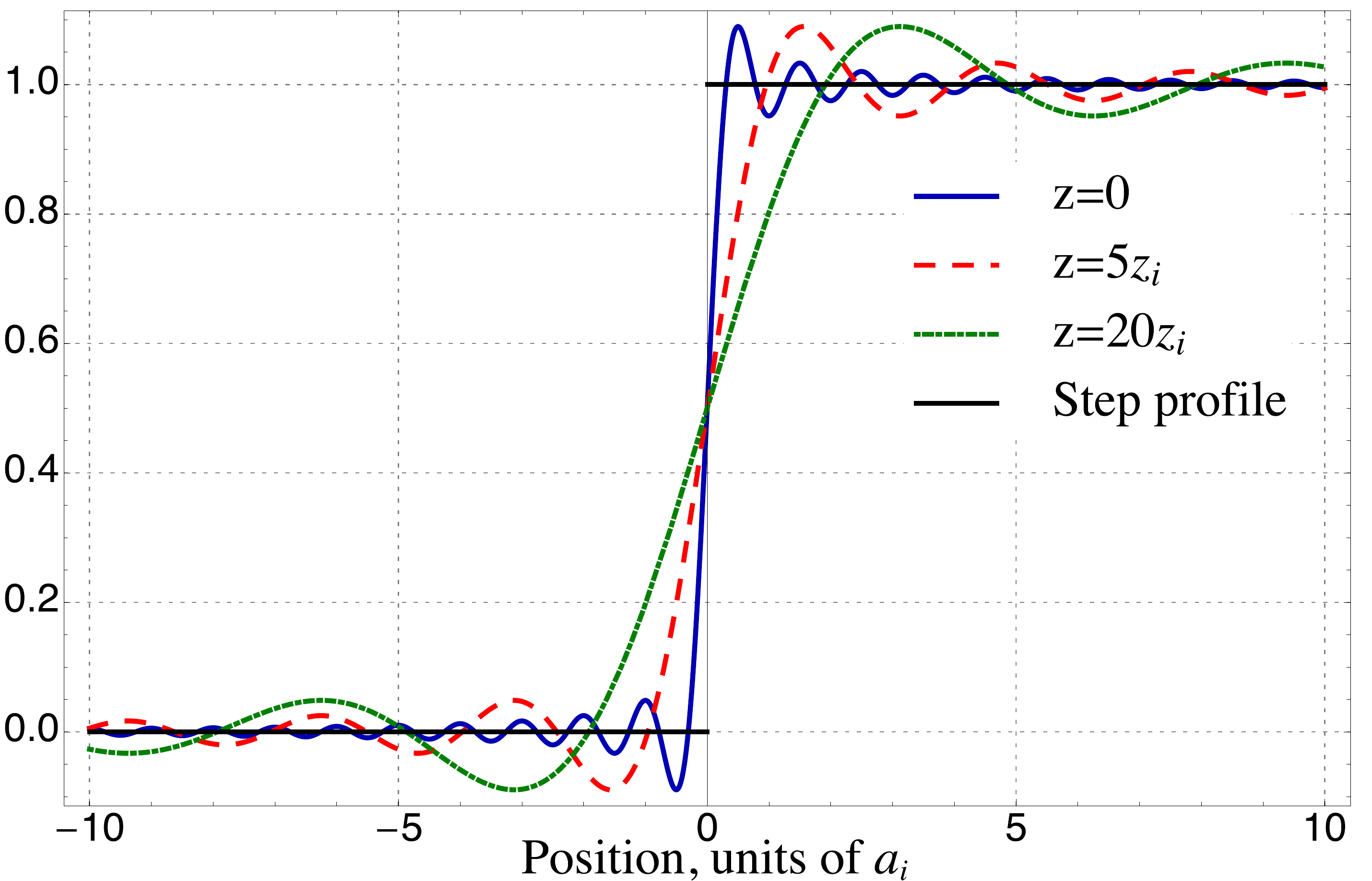}
     \caption{Phase profiles $\varphi(x)/\varphi_0$ of a step~(\ref{eq:phase_step}), depicted by in solid black, reconstructed
according to Eq.~(\ref{eq:phase_reconstruction_z}): the solid blue line
corresponds to $z=z_i$, dashed red line corresponds to $z=5z_i$. Lateral
position is in units of $a_i$.}
\label{fig:PhaseProfileStep_z}
\end{figure}

We illustrate reconstruction~(\ref{eq:phase_reconstruction_z}) of a defocused
phase step~(\ref{eq:phase_step}) by a system with matched numerical apertures
${\rm NA_i}={\rm NA_d}$. Intensities detected in the first and second
quadrants are obtained using the transfer
functions~(\ref{eq:PhaseTFReduced_xy}). The characteristic axial length scale
of defocusing is the depth of focus $z_i=\lambda/(2{\rm NA_i^2})$. Intensities
calculated at $z=z_i$ and $z=5z_i$ are shown in
Fig.~\ref{fig:IntensityStepQ12}. From these plots we can unambiguously
identify the defocus positions. Indeed, pairs of intensities calculated at the
axial positions $z=z_i$ and $z=5z_i$ are shifted with respect to one another
by distances $\delta x= 2 a_i$ and $\delta x= 10 a_i$, respectively. Phase
profiles reconstructed according to Eq.~(\ref{eq:phase_reconstruction_z}) are
demonstrated in Fig.~\ref{fig:PhaseProfileStep_z}. We choose an arbitrary
constant so that $\varphi(0)=\varphi_0/2$. The reconstructed profile
$\varphi(x)/\varphi_0=1/2+(1/\pi) {\rm Si}(2\pi x/a_i)$ calculated at $z=z_i$
is identical to the in-focus one. One can show that detection-limited phase
reconstruction is possible for relatively small defocus $|z|<2z_i$. In
contrast, the profile reconstructed at $z=5z_i$ is characterized by a reduced
spatial resolution. The reason is that at this relatively large defocus the
transfer function $\tilde P^x_\theta(k_x,z)=0$ at $|k_x|=k_c=0.32/a_i$ within
the detection bandwidth $|k_{x}|\le 1/a_i$. To avoid the singularities
in~(\ref{eq:phase_reconstruction_z}), we thus restrict the bandwidth of
integration to a smaller range $|k_x|< k_c$. The resulting profile is given by
$\varphi(x)/\varphi_0=1/2+(1/\pi) {\rm Si}(2\pi x k_c)$. Oscillations observed
around discontinuity at $x=0$ are due to the limited bandwidth of integration
in~(\ref{eq:phase_reconstruction_z}). As the defocus distance is increased,
the integration bandwidth is gradually reduced. We find that even at a
significant defocus distance $z=20z_i$, the first zero of the tilt transfer
function occurs at $k_c=0.16/a_i$, so that the bandwidth is reduced only two
times compared to $z=5z_i$.

\section{Conclusion}

In this work we considered quantitative phase imaging using a partitioned
detection aperture microscope introduced in Ref.~\cite{Mertz2012}. Our
single-exposure method is characterized by a simple optical design that is
light efficient and insensitive to polarization. The microscope may be
implemented in transmission or reflection modes~\cite{Mertz2012,Barankov2013}.
Within the wave-optic paraxial approximation we derived the three-dimensional
spread functions of intensity and phase~(\ref{eq:PSF_0})
and~(\ref{eq:PSF_small_phase}), analyzed their properties, and refined the
phase-imaging approach of Ref.~\cite{Mertz2012,Barankov2013}, making it
suitable for imaging smooth and rough samples. The formalism is summarized in
Eq.~(\ref{eq:phase_reconstruction_precise}). In particular, we demonstrated
that the spatial resolution of our method is limited by the detection aperture
in contrast to Refs.~\cite{Mertz2012,Barankov2013} where the resolution was
defined by a smaller tilt range. We also showed how our method
reconstructs the profile of a phase step provided is optical path-length is
smaller than the wavelength. Further, we extended the method to imaging of
defocused phase objects, as described in Eqs.~(\ref{eq:Intensities_sheared}),
(\ref{eq:Tilts_z}) and~(\ref{eq:phase_reconstruction_z}). Even for relatively
large object defocus, the method provides precise reconstructions of rough and
smooth samples with spatial resolution depending on the defocus distance.

\section{Acknowledgments} 
Financial support for this work was partially provided by NIH CA182939.

\section{Appendix}
We provide calculations of spread functions, which, for the sake of clarity, were omitted in the main text.

In our calculations we employ the following convention for Fourier transforms
\bea
f(\vec  r)=\int {\rm d^2} \vec k\, e^{i2\pi \vec k{\vec  r}}\tilde f(\vec k),\quad
\tilde f(\vec k)=\int {\rm d^2} {\vec  r}\, e^{-i2\pi \vec k{\vec  r}}f({\vec  r}).
\eea

\subsection{Phase-gradient expansion of transfer functions}
In the limit of small wavevectors $\vec k\to 0$, one can simplify~(\ref{eq:PSF_small_phase}). Assuming that the aperture is a real function, ${\rm A^*_c}={\rm A_c}$, we expand the difference of the functions into the Taylor series and keep the first non-vanishing term 
\be
{\rm A_c}\lp-\frac{f_c}{k}\vec q_+\rp-{\rm A^*_c}\lp-\frac{f_c}{k}\vec q_-\rp\approx \vec k\,\nabla_{\vec q}{\rm A_c}\lp-\frac{f_c}{k}\vec q\rp,
\ee
where $\vec q_\pm=\vec q\pm \vec k/2$. Substituting this expression in Eq.~(\ref{eq:PSF_small_phase}), and restricting the expansions to terms linear in $\vec k$, we derive an approximate phase transfer function
\bea\label{eq:PSF_phase_smooth_small}
\tilde P_\varphi (\vec k,z)&\approx & -\frac{i}{f_c\Omega_c k}\int {\rm d^2}\vec r' \left|{\rm A_d}\lp \frac{f_e}{f_c}\vec r' -\vec b\rp\right|^2\nn
&&\times\,\vec k\nabla{\rm A_c}\lp-\vec r'\rp,
\eea
where we introduced the integration variable $\vec r=(f_c/k)\vec q$. The Fourier components of the function are most significant along the directions of rapid variations of the illumination aperture. For a uniformly illuminated aperture, the directions are orthogonal to the aperture edge.

The intensity transfer function~(\ref{eq:PSF_0}) is approximated by its value at $\vec k=0$ given by
\be\label{eq:PSF_0_smooth_small}
\tilde P_\alpha(\vec k=0,z)=\frac{2}{f^2_c\Omega_c}\int {\rm d^2}\vec r'\, \left|{\rm A_d}\lp\frac{f_e}{f_c} \vec r' -\vec b\rp\right|^2 {\rm A_c}\lp-\vec r'\rp.
\ee
The next non-vanishing order of $\tilde P_\alpha(\vec k,z)$ scales as $\vec k^2$ and can be neglected, since we limit the expansions to terms linear in $\vec k$.

Substituting the transfer functions~(\ref{eq:PSF_phase_smooth_small}) and
(\ref{eq:PSF_0_smooth_small}) in Eq.~(\ref{eq:intensity_phase_fourier}), and
performing the inverse Fourier transform, we obtain the intensity at the
detector plane~(\ref{eq:intensity_image_smooth_small}).

\subsection{Spread functions for intensity and phase}
Below we define the spread functions of phase, tilt angles and intensity
when the detection aperture is a square. We assume that ${\rm NA_d}\ge {\rm
NA_i}$. In the following, to simplify the notations we use the dimensionless coordinates
\be
z/z_i\to z,\quad \vec r/a_i\to \vec r,
\ee
where 
$a_i=\lambda/(2{\rm NA_i})$, and $z_i=\lambda/(2{\rm NA^2_i})$
define the lateral and axial spatial scales, identified as the lateral resolution and the depth of focus of the illumination aperture, respectively. Wavevectors are also rescaled: $\vec k a_i\to \vec k$.

Calculation of the spread functions is relatively straightforward and somewhat tedious, and we provide the results without detailed derivations.

\subsection{The intensity spread function}

\begin{figure}[htbp]
	\centering
    \includegraphics[width=8cm]{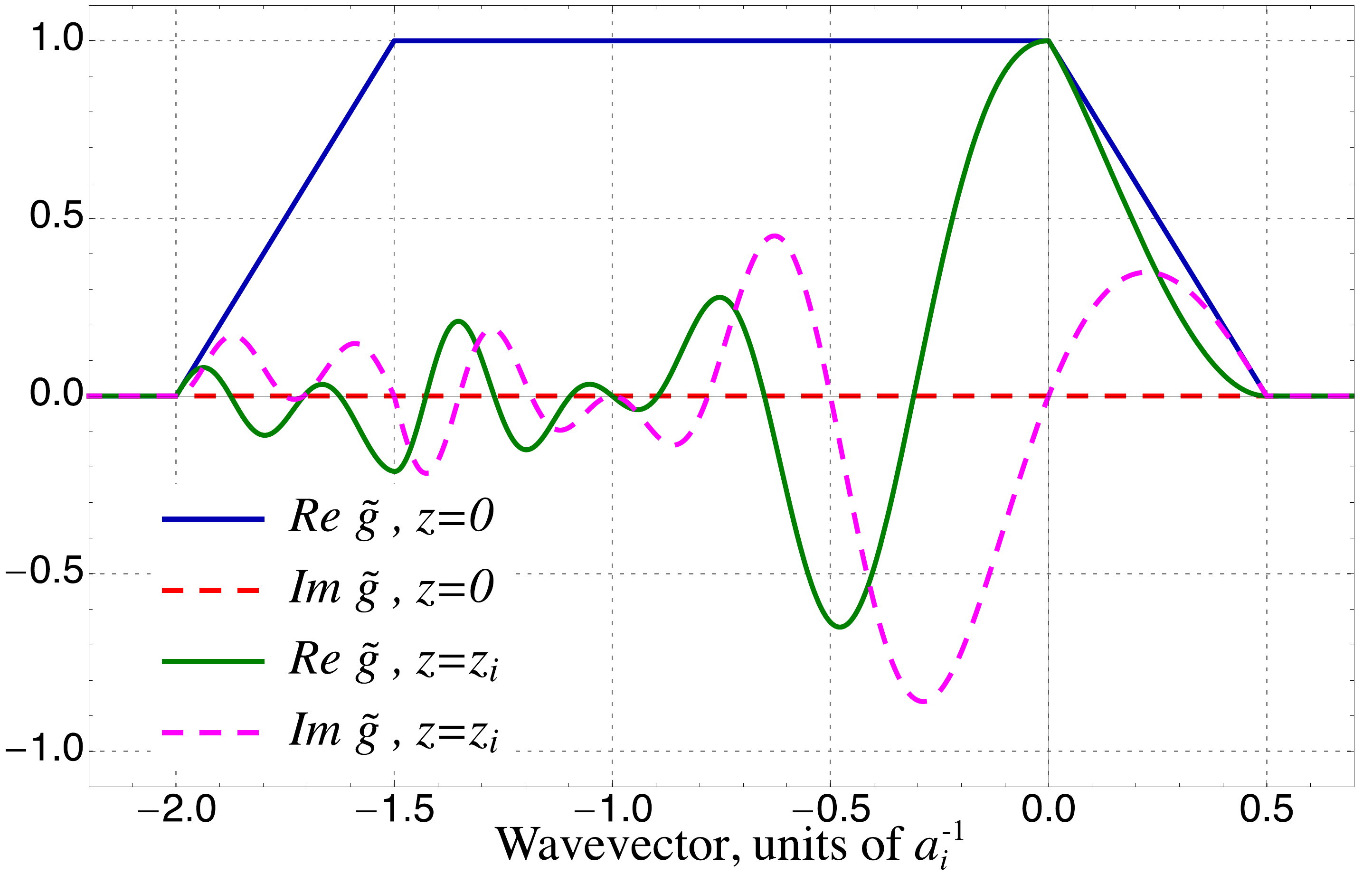}
     \caption{Plots of the function $\tilde g(q,z)$ at different defocus distances and in the case ${\rm NA_d}=2{\rm NA_i}$. Solid blue and red dashed lines depict the real and imaginary parts of $\tilde g(q,z)$, respectively, calculated at $z=0$; solid green and dashed magenta lines depict the real and imaginary parts of $\tilde g(q,z)$ calculated at $z=z_i$.}
\label{fig:g_function}
\end{figure}

Due to the square symmetry of the illumination aperture, the intensity transfer functions $P^{(m)}_\alpha(\vec r)$~(\ref{eq:PSF_0}), where $m$ identifies the quadrant of the detection lens, are given by 
\be\label{eq:IntensitySFFourierZ}
\tilde P^{(m)}_\alpha=\frac{1}{4}\left\{\begin{array}{ll}
\tilde g(k_x,z) \tilde g(k_y,z)+\tilde g^*(-k_x,z)\tilde g^*(-k_y,z), &  \mbox{m=1,}\\
\tilde g(-k_x,z)\tilde g(k_y,z)+\tilde g^*(k_x,z)\tilde g^*(-k_y,z), &  \mbox{m=2,}\\
\tilde g(-k_x,z)\tilde g(-k_y,z)+\tilde g^*(k_x,z)\tilde g^*(k_y,z), &  \mbox{m=3,}\\
\tilde g(k_x,z)\tilde g(-k_y,z)+\tilde g^*(-k_x,z)\tilde g^*(k_y,z), &  \mbox{m=4},
\end{array}\right.
\ee
where where we defined a piece-wise complex-valued function
\be\label{eq:gfunction_fourier_z}
\tilde g(q,z)=\frac{e^{-i2\pi zq^2}}{i2\pi z q}\left\{\begin{array}{ll}
 e^{i2\pi z q}-e^{i4\pi z q^2}, &  \mbox{$0\le q< 1/2$},\\
 e^{i2\pi z q}-1, &  \mbox{$1/2-\beta\le q < 0$,}\\
 e^{i4\pi z q (q+\beta)}-1, &  \mbox{$-\beta\le q< 1/2 -\beta$,}\\
0, &  \mbox{$q< -\beta$, or $q\ge 1/2$},
\end{array}\right.
\ee
the parameter $\beta={\rm NA_d}/{\rm NA_i}\ge 1$, and $\tilde g^*(q,z)$ is the complex conjugate of $\tilde g(q,z)$. The real and imaginary parts of $\tilde g(q,z)$ are plotted in Fig.~\ref{fig:g_function} when ${\rm NA_d}=2{\rm NA_i}$ for in-focus and out-of-focus imaging at $z=z_i$.

The real-space representation of this function is simplified at the in-focus position $z=0$
\be\label{eq:g_zero}
g(x,0)=\beta\, e^{-i\pi x(\beta -1/2)}{\rm sinc}\lp \pi x/2\rp{\rm sinc}\lp \pi x\beta\rp.
\ee

The in-focus ($z=0$) intensity spread functions are given by
\be
P^{(m)}_\alpha(\vec r)=\rho(\vec r)\left\{\begin{array}{ll}
\cos\lb\pi(x+y)(\beta-1/2)\rb, &  m=1,\\
\cos\lb\pi(-x+y)(\beta-1/2)\rb, &  m=2,\\
\cos\lb\pi(-x-y)(\beta-1/2)\rb, &  m=3,\\
\cos\lb\pi(x-y)(\beta-1/2)\rb, &  m=4,
\end{array}\right.
\ee
where $\rho(\vec r)=(\beta^2/2){\rm sinc}(\pi x/2){\rm sinc}(\pi y/2){\rm
sinc}(\pi x\beta){\rm sinc}(\pi y\beta)$. The intensity spread functions
calculated at the in-focus position $z = 0$ for different quadrants are
plotted in Fig.~\ref{fig:PhaseIntensitySFzero}.

\subsection{The spread functions for phase and light tilts}
Due to the square symmetry of the detection and illumination apertures, the phase transfer functions $\tilde P^{(m)}_\varphi(\vec k,z)$~(\ref{eq:PSF_small_phase}), where $m$ identifies the quadrant of the lens, are  given by
\be\label{eq:PhaseSFFourierZ}
\tilde P^{(m)}_\varphi=-\frac{i}{4}\left\{\begin{array}{ll}
\tilde g(k_x,z) \tilde g(k_y,z)-\tilde g^*(-k_x,z)\tilde g^*(-k_y,z), &  \mbox{m=1,}\\
\tilde g(-k_x,z)\tilde g(k_y,z)-\tilde g^*(k_x,z)\tilde g^*(-k_y,z), &  \mbox{m=2,}\\
\tilde g(-k_x,z)\tilde g(-k_y,z)-\tilde g^*(k_x,z)\tilde g^*(k_y,z), &  \mbox{m=3,}\\
\tilde g(k_x,z)\tilde g(-k_y,z)-\tilde g^*(-k_x,z)\tilde g^*(k_y,z), &  \mbox{m=4},
\end{array}\right.
\ee
where the function $\tilde g(q,z)$ is defined in Eq.~(\ref{eq:gfunction_fourier_z}).

The in-focus phase spread functions are thus given by
\be
P^{(m)}_\varphi=-\rho(\vec r)\left\{\begin{array}{ll}
\sin\lb\pi(x+y)(\beta-1/2)\rb, &  m=1,\\
\sin\lb\pi(-x+y)(\beta-1/2)\rb, &  m=2,\\
\sin\lb\pi(-x-y)(\beta-1/2)\rb, &  m=3,\\
\sin\lb\pi(x-y)(\beta-1/2)\rb, &  m=4,
\end{array}\right.
\ee
where $\rho(\vec r)=(\beta^2/2){\rm sinc}(\pi x/2){\rm sinc}(\pi y/2){\rm sinc}(\pi x\beta){\rm sinc}(\pi y\beta)$. The phase spread functions at the in-focus position $z=0$ for different quadrants are plotted in Fig.~\ref{fig:PhaseIntensitySFzero}.

The in-focus tilt transfer functions are calculated as linear combinations of the phase spread functions according to Eq.~(\ref{eq:tilts})
\bea\label{eq:TiltsSFFourierInFocus}
\tilde P^x_\theta &=& \tilde P_\varphi^{(1)}+\tilde P_\varphi^{(4)}-\tilde P_\varphi^{(2)}-\tilde P_\varphi^{(3)},\nn
\tilde P^y_\theta &=& \tilde P_\varphi^{(1)}+\tilde P_\varphi^{(2)}-\tilde P_\varphi^{(3)}-\tilde P_\varphi^{(4)}.
\eea
Taking the limit $z\to 0$ in Eqs.~(\ref{eq:PhaseSFFourierZ}) and substituting the result into~(\ref{eq:TiltsSFFourierInFocus}), we obtain
\bea
\tilde P^x_\theta(\vec k)=2i\, \tilde g_-(k_x)\tilde g_+(k_y),\quad
\tilde P^y_\theta(\vec k)=2i\, \tilde g_-(k_y)\tilde g_+(k_x),
\eea
where
\bea\label{eq:gminusplus_zero}
\tilde g_-(q)&=&\lb \tilde g(-q,0)-\tilde g(q,0)\rb/2,\nn
\tilde g_+(q)&=&\lb \tilde g(-q,0)+\tilde g(q,0)\rb/2.
\eea
The functions $\tilde g_-(q)$ and $\tilde g_+(q)$ are the odd and even functions, respectively. For positive arguments, the functions are
\be
\tilde g_-(q)=\left\{\begin{array}{ll}
q, &  \mbox{if $0\le q < 1/2$,}\\
1/2, &  \mbox{$1/2 \le q < \beta-1/2$,}\\
\beta -q, &  \mbox{$\beta-1/2 \le q < \beta$,}\\
0, &  \mbox{$q \ge \beta$,}
\end{array}\right.
\ee
\be
\tilde g_+(q)=\left\{\begin{array}{ll}
1-q, &  \mbox{$0\le q < 1/2$,}\\
1/2, &  \mbox{$1/2 \le q < \beta-1/2$,}\\
\beta -q, &  \mbox{$\beta-1/2 \le q < \beta$,}\\
0, &  \mbox{$q \ge \beta$.}
\end{array}\right.
\ee
The functions are shown in Fig.~\ref{fig:TiltSF} for the case $\beta=2$, that is when ${\rm NA_d}=2{\rm NA_i}$.

Real-space representations of the functions are given by
\bea
g_+(x)= {\rm Re}\, g(x,0),\quad g_-(x)=-i\,{\rm Im}\, g(x,0),
\eea
where ${\rm Re}$ and ${\rm Im}$ are the real and imaginary parts,
respectively, and the function $g(x,0)$ is defined in Eq.~(\ref{eq:g_zero}).

The tilt spread functions are the products of the functions
\bea
P^x_\theta(\vec r)= 2i\, g_-(x)g_+(y),\quad P^x_\theta(\vec r)= 2i\, g_-(y)g_+(x).
\eea
The functions are plotted in Fig.~\ref{fig:TiltsSpreadSpace} in the case $\beta=2$, that is when ${\rm NA_d}=2{\rm NA_i}$.

\subsection{Properties of the 3D phase spread functions}

At a defocus distance $z$, the phase spread functions exhibit lateral shearing
and blurring. These effects can be traced to the behavior of the corresponding
transfer functions $\tilde g(q,z)$ in~(\ref{eq:gfunction_fourier_z}) at
different spatial frequencies. For example, within the tilt range
$|k_{x,y}|\le 1/2$ (in dimensionless units), the transfer functions are
proportional to the phase factor encoding lateral shearing in real space
$\tilde P^{(m)}_\varphi(\vec k,z)\sim \exp\lb i2\pi (\pm k_x\pm k_y)z\rb$,
where the signs depend on the quadrant $m$.

Lateral shearing of the spread functions allows identification of the
corresponding defocus position $z$. For simplicity, let us consider the
reduced transfer functions in which one of the two components of the
wavevector is set to zero. Depending on the value of the remaining second
component, the transfer functions are related to one another. For the first
and second quadrants we find
\be\label{eq:RelPhaseTFReduced_x}
\tilde P_\varphi^{(1)}(k_x,0,z)=-\tilde P_\varphi^{(2)}(k_x,0,z)h(k_x,z),
\ee
where
\be
h(q,z)=\left\{\begin{array}{ll}
e^{i4\pi z q }, &  \mbox{$|q| < 1/2$,}\\
e^{i2\pi z q (1-2|q|) }, &  \mbox{$1/2 \le |q| < \beta-1/2$,}\\
e^{i4\pi z \beta q} , &  \mbox{$\beta-1/2 \le |q| < \beta$.}
\end{array}\right.
\ee
Certainly, $\tilde P_\varphi^{(1)}(0,k_y,z)=\tilde P_\varphi^{(2)}(0,k_y,z)$.

The reduced transfer functions of the first and fourth quadrants are also
proportional to one another
\be\label{eq:RelPhaseTFReduced_y}
\tilde P_\varphi^{(1)}(0,k_y,z)=-\tilde P_\varphi^{(4)}(0,k_y,z)h(k_y,z).
\ee
We also obtain $\tilde P_\varphi^{(1)}(k_x,0,z)=\tilde
P_\varphi^{(4)}(k_x,0,z)$. Similar relations can be derived for other pairs of
the transfer functions.

\subsection{3D spread functions when ${\rm NA_i}={\rm NA_d}$}
\label{section:matched out-of-focus imaging}

The case of matching numerical apertures ${\rm NA_i}={\rm NA_d}$ is special,
since, according to Eqs.~(\ref{eq:RelPhaseTFReduced_x})
and~(\ref{eq:RelPhaseTFReduced_y}), the pairs of reduced transfer functions
are related to one another by the phase factors $e^{i4\pi z k_x}$ and
$e^{i4\pi z k_y}$, respectively, for all frequencies within the detection
bandwidth $|k_{x,y}|\le 1$ (in dimensionless units).

Similar to Eqs.~(\ref{eq:gminusplus_zero}), we introduce odd and even functions of the argument $q$ within the bandwidth $|q|\le 1$:
\be\label{eq:gminus_z}
\tilde g_-(q,z)=\left\{\begin{array}{ll}
q\,{\rm sinc}\lb 2\pi z q^2\rb , &  |q|\le 1/2\\
\frac{\sin\lb 2\pi z |q| (1-|q|)\rb}{2\pi zq}, &  1/2<|q|\le 1,
\end{array}\right.
\ee
\be
\tilde g_+(q,z)=(1-|q|)\,{\rm sinc}\lb 2\pi z q (1-|q|)\rb, \quad  |q|\le 1.
\ee
The functions $\tilde g_-(q,z)$ and $\tilde g_+(q,z)$ equal zero at $|q|> 1$.

The reduced phase transfer functions calculated for different quadrants are given by  
\be\label{eq:PhaseTFReduced_xy}
\tilde P_\varphi^{(m)}(k_x,0,z)=\left\{\begin{array}{ll}
\tilde f(k_x,z), &  m=1,\\
\tilde f^*(k_x,z), &  m=2,\\
\tilde f^*(k_x,z), &  m=3,\\
\tilde f(k_x,z), &  m=4,
\end{array}\right.
\ee
\be\label{eq:PhaseTFReduced_yx}
\tilde P_\varphi^{(m)}(0,k_y,z)=
\left\{\begin{array}{ll}
\tilde f(k_y,z), &  m=1,\\
\tilde f(k_y,z), &  m=2,\\
\tilde f^*(k_y,z), &  m=3,\\
\tilde f^*(k_y,z), &  m=4,
\end{array}\right.
\ee
where
\be
\tilde f(q,z)=\frac{i}{2}\,e^{i2\pi z q}\tilde g_-(q,z).
\ee

\begin{figure}[htbp]
	\centering
    \includegraphics[width=8cm]{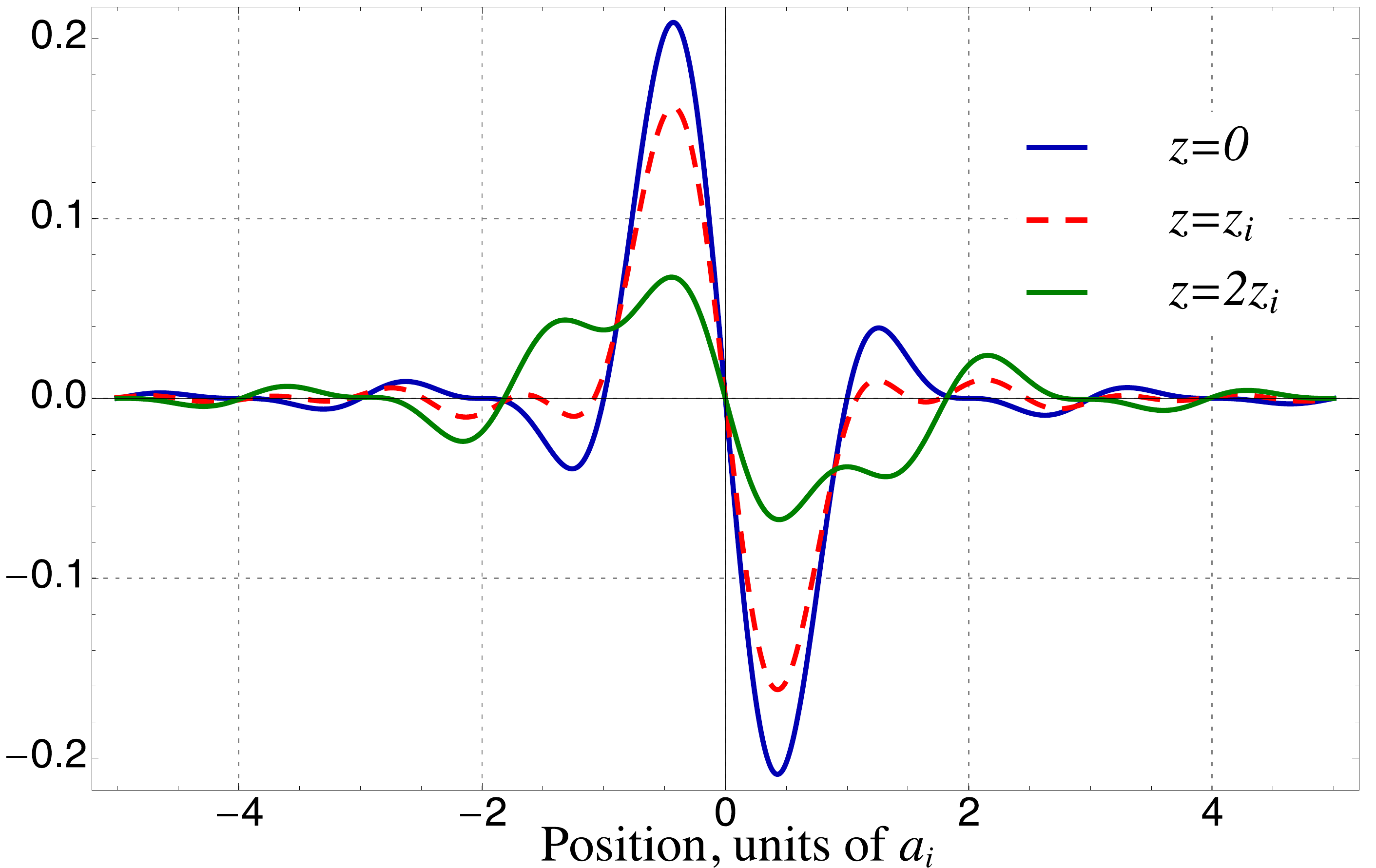}
     \caption{Plots of ${\cal P}_\varphi (u,z)$ for different values of $z$. The lateral position is in units of $a_i$}
\label{fig:PhaseSFAve}
\end{figure}

The averaged spread functions $\la P_\varphi^{(m)}(x,y,z)\ra_y$ and $\la P_\varphi^{(m)}(x,y,z)\ra_x$ are expressed in terms of the function ${\cal P}(x,z)$
\be\label{eq:PhaseSFReduced_x}
\la P_\varphi^{(m)}(x,y,z)\ra_y=
\left\{\begin{array}{ll}
{\cal P}_\varphi(x+z,z), &  m=1,\\
{\cal P}_\varphi(-x+z,z), &  m=2,\\
{\cal P}_\varphi(-x+z,z), &  m=3,\\
{\cal P}_\varphi(x+z,z), &  m=4,
\end{array}\right.
\ee 
\be\label{eq:PhaseSFReduced_y}
\la P_\varphi^{(m)}(x,y,z)\ra_x=
\left\{\begin{array}{ll}
{\cal P}_\varphi(y+z,z), &  m=1,\\
{\cal P}_\varphi(y+z,z), &  m=2,\\
{\cal P}_\varphi(-y+z,z), &  m=3,\\
{\cal P}_\varphi(-y+z,z), &  m=4,
\end{array}\right.
\ee
where $\la..\ra_{x,y}$ is the average over $x$ or $y$ coordinates, and we defined
\be
{\cal P}_\varphi(u,z)=-\int_0^{1}{\rm d}q\,\sin\lp 2\pi u q\rp \tilde g_-(q,z).
\ee

The dependence of function ${\cal P}_\varphi(u,z)$ on the coordinate $u$ is shown in Fig.~\ref{fig:PhaseSFAve} for different defocus distances.

Eqs.~(\ref{eq:RelPhaseTFReduced_x}) and~(\ref{eq:RelPhaseTFReduced_y}) can be
used to identify $z$. Once the defocus distance $z$ is known, we can separate
the effects of blurring and shearing. The latter is accounted for by including
the corresponding phase factors in the definitions of the tilt transfer
functions:
\bea\label{eq:TiltsSFSheared}
\tilde P^x_\theta &=& e^{-i2\pi z(k_x+k_y)}\tilde P_\varphi^{(1)}+e^{-i2\pi z(k_x-k_y)}\tilde P_\varphi^{(4)}\nn
&&-e^{i2\pi z(k_x-k_y)}\tilde P_\varphi^{(2)}-e^{i2\pi z(k_x+k_y)}\tilde P_\varphi^{(3)},\nn
\tilde P^y_\theta &=& e^{-i2\pi z(k_x+k_y)}\tilde P_\varphi^{(1)}+e^{i2\pi z(k_x-k_y)}\tilde P_\varphi^{(2)}\nn
&&-e^{i2\pi z(k_x+k_y)}\tilde P_\varphi^{(3)}-e^{-i2\pi z(k_x-k_y)}\tilde P_\varphi^{(4)}.
\eea

The tilt transfer functions are given by 
\bea\label{eq:TFTiltsMatchedApertures}
\tilde P^x_\theta(\vec k,z)&=& 2i\,\tilde g_-(k_x,z) \tilde g_+(k_y,z),\nn
\tilde P^y_\theta(\vec k,z)&=& 2i\,\tilde g_-(k_y,z) \tilde g_+(k_x,z),
\eea
and can be used for phase reconstruction~(\ref{eq:phase_reconstruction_z}) with the resolution defined by the detection aperture, at least for small enough defocus distances.

\subsection{Imaging an out-of-focus phase step}
A phase step located a distance $z$ away from the focal plane is characterized in frequency space by~(\ref{eq:phase_step_fourier}). 

For a system with matched illumination and detection apertures, ${\rm NA_i}={\rm NA_d}$, the intensities detected in the first and second quadrants are obtained using the transfer functions~(\ref{eq:gminus_z}):
\bea\label{eq:IntensitiesStepFullAperture}
I^{(1)}_\varphi&=&\frac{I_0\varphi_0}{2\pi}\int_0^{1}{\rm d}q\,\frac{\tilde g_-(q,z)}{q} \cos\lb 2\pi q \lp x+z\rp\rb,\nn
I^{(2)}_\varphi&=&-\frac{I_0\varphi_0}{2\pi}\int_0^{1}{\rm d}q\,\frac{\tilde g_-(q,z)}{q} \cos\lb 2\pi q \lp x-z\rp\rb.
\eea
Fig.~\ref{fig:IntensityStepQ12} illustrates these functions for defocus distances $z=0$, $z=5$ and $z=20$ (in dimensionless units). 

\subsection{Out-of focus imaging when ${\rm NA_i}<{\rm NA_d}$}
\label{section:unbalanced out-of-focus imaging}

When the numerical apertures of the illumination and detection are not
matched, then, according to Eqs.~(\ref{eq:RelPhaseTFReduced_x})
and~(\ref{eq:RelPhaseTFReduced_y}), the pairs of reduced transfer functions
are related to one another by the phase factors $e^{i4\pi z k_x}$ and
$e^{i4\pi z k_y}$ only within the tilt bandwidth $|k_{x,y}|\le 1/2$ (in
dimensionless units). The phase factor relating the functions changes for
spatial frequencies outside this range. In that case, phase reconstruction of
out-of-focus objects remains possible, although with a spatial resolution
reduced compared to the case of matched numerical apertures, at least for
small enough defocus distance $z$.

The formalism presented in Section~\ref{section:matched out-of-focus imaging}
remains applicable provided the bandwidth of the transfer
functions~(\ref{eq:PhaseTFReduced_xy}), (\ref{eq:PhaseTFReduced_yx}) and
(\ref{eq:TFTiltsMatchedApertures}) is reduced to the tilt range
$|k_{x,y}|\le 1/2$ (in dimensionless units). The transfer functions can be
used for identification of defocus distance and phase
reconstruction~(\ref{eq:phase_reconstruction_z}) with the resolution defined
by the tilt range for relatively small defocus distances.


\end{document}